\documentclass[aps,prl,reprint,superscriptaddress]{revtex4-1}

\usepackage{graphicx}
\usepackage{epsfig}
\usepackage{amsmath,bbm,amssymb,bm,times}
\usepackage[colorlinks,linkcolor=blue,citecolor=blue]{hyperref}

\usepackage{stmaryrd}
\usepackage[normalem]{ulem}
\usepackage{comment}
\usepackage[usenames,dvipsnames]{color}
  
\definecolor{dblue}{rgb}{0.0, 0.0, 0.6}

\definecolor{dgreen}{rgb}{0.0, 0.5, 0.0}

\newcommand{\eqn}[1]{
\begin{eqnarray}
	#1
\end{eqnarray}
}

\begin{document}

\title{Anomalous Topological Active Matter}

\author{Kazuki Sone}
\email{sone@noneq.t.u-tokyo.ac.jp}
\affiliation{Department of Applied Physics, The University of Tokyo, 7-3-1 Hongo, Bunkyo-ku, Tokyo 113-8656, Japan}
\author{Yuto Ashida}
\email{ashida@ap.t.u-tokyo.ac.jp}
\affiliation{Department of Applied Physics, The University of Tokyo, 7-3-1 Hongo, Bunkyo-ku, Tokyo 113-8656, Japan}
\affiliation{Department of Physics, University of Tokyo,  7-3-1 Hongo, Bunkyo-ku, Tokyo 113-0033, Japan}

\date{\today}

\begin{abstract}
Active systems exhibit spontaneous flows induced by self-propulsion of microscopic constituents and can reach a nonequilibrium steady state without an external drive. Constructing the analogy between the quantum anomalous Hall insulators and active matter with spontaneous flows, we show that topologically protected sound modes can arise in a steady-state active system in continuum space. We point out that the net vorticity of the steady-state flow, which acts as a counterpart of the gauge field in condensed-matter settings, must vanish under realistic conditions for active systems. The quantum anomalous Hall effect thus provides design principles for realizing topological metamaterials. We propose and analyze the concrete minimal model and numerically calculate its band structure and eigenvectors, demonstrating the emergence of nonzero bulk topological invariants with the corresponding edge sound modes. This new type of topological active systems can potentially expand possibilities for their experimental realizations and may have broad applications to practical active metamaterials. Possible realization of non-Hermitian topological phenomena in active systems is also discussed.
\end{abstract}

\maketitle

Topologically nontrivial bands, which have been at the forefront of condensed matter physics \cite{Thouless1982,Simon1983,Haldane1988,Kane2005,Hasan2010,Qi2011}, can also appear  in various classical systems such as photonic \cite{Haldane2008,Lu2014,Rechtsman2013,Khanikaev2013} and phononic systems \cite{Huber2016,Nash2015,Susstrunk2015,Kane2014,Yang2015,Fleury2016,Delplace2017}. Such topologically nontrivial systems exhibit unidirectional modes that propagate along the edge of a sample and are immune to disorder. The existence of  edge modes originates from the nontrivial topology characterized by bulk topological invariants of underlying photonic or acoustic band structures. The topological edge modes give rise to novel functionalities potentially applicable to, e.g.,  sonar detection and heat diodes \cite{Huber2016,Yang2015}. Furthermore, they are argued to be closely related to the mechanism of  robustness in biological systems \cite{Murugan2017,Dasbiswas2017}.

On another front, active matter, a collection of self-driven particles, has attracted much interest as an ideal platform to study biological physics  \cite{Brugues2014,Saw2017,Kawaguchi2017} and out-of-equilibrium statistical physics  \cite{Ganguly2013,Fodor2016,Krishnamurthy2016,Mandal2017,Pietzonka2018,Martin2018}. While a prototype of active matter has been originally introduced to understand animal flocking behavior \cite{Vicsek1995,Vicsek2012}, recent experimental developments have allowed one to manipulate and observe artificial active systems in a controlled manner by utilizing Janus particles \cite{Jiang2010}, catalytic colloids \cite{Palacci2013} and external feedback control \cite{Khadka2018}.

The aim of this Letter is to show that a topologically nontrivial feature can ubiquitously emerge in a nonequilibrium steady state of active matter and demonstrate it by analyzing the concrete minimal model, which can be realized with current experimental techniques. Specifically, we first point out that the net vorticity of the steady-state flow must vanish under realistic conditions for active systems in the continuum space. Since the vorticity in active matter can act as a counterpart of the magnetic field in condensed matter systems, this fact  indicates that the quantum anomalous Hall effect (QAHE) naturally provides design principles for realizing topological metamaterials. We propose and analyze the active matter model inspired by the flat-band ferromagnet featuring the QAHE \cite{Ohgushi2000}. We numerically calculate its band structure and eigenvectors, and demonstrate that they exhibit nonzero topological invariants with the corresponding edge modes.  Possible relation to non-Hermitian topological phenomena is also discussed.

Topological edge modes of active matter have been recently discussed by several authors \cite{Souslov2017,Shankar2017,Souslov2018,Dasbiswas2017}. There, the presence of nonzero net vorticity of the active flows, which can act as an effective magnetic field, was crucial to support topological edge modes reminiscent of the quantum Hall effect \cite{Klitzing1980,Thouless1982}. Yet, this required the introduction of rather intricate structures in active systems such as large defects \cite{Souslov2017}, curved surface \cite{Shankar2017}, and rotational forces \cite{Dasbiswas2017,Souslov2018}. 
One of the novel aspects introduced by this Letter is to eliminate these  bottlenecks by constructing the analogy to the QAHE, significantly expanding possibilities for realizing topological metamaterials. Our proposal is based on the simplest setup on a flat continuum space with assuming no internal degrees of freedom of active particles. This class of systems is directly relevant to many realistic setups of active systems \cite{Sokolov2007,Cavagna2010,Deseigne2010,Schaller2010,Nishiguchi2018} and  our design principle is applicable beyond the minimal model proposed here.

{\it Emergent effective Hamiltonian for active matter.---} To describe collective dynamics of active matter, we use the Toner-Tu equations \cite{Toner1995,Toner1998,Toner2005,Marchetti2013}, which are the hydrodynamic equations for active matter with a polar-type interaction: 
\eqn{
\partial_t \rho + \nabla \cdot (\rho \mathbf{v}) = 0,
\label{toner-tu-eq1}
}
\eqn{
&{}& \partial_t \mathbf{v} + \lambda (\mathbf{v} \cdot \nabla) \mathbf{v} + \lambda_2 (\nabla \cdot \mathbf{v}) \mathbf{v} + \lambda_3  \nabla |\mathbf{v}|^2 \nonumber\\
&=& (\alpha-\beta|\mathbf{v}|^2)\mathbf{v}-\nabla P \nonumber\\
&&+D_B \nabla (\nabla \cdot \mathbf{v}) + D_T \nabla^2 \mathbf{v} + D_2(\mathbf{v} \cdot \nabla)^2 \mathbf{v} + \mathbf{f},
\label{toner-tu-eq2}
}
where $\rho({\bf r},t)$ is the density field of active matter and $\mathbf{v}({\bf r},t)$ is the local average of velocities of self-propelled particles. Equation~\eqref{toner-tu-eq1} presents the equation of continuity. In Eq.~\eqref{toner-tu-eq2}, the first term of its right-hand side suggests a preference for a nonzero constant speed $|\mathbf{v}|=\sqrt{\alpha/\beta}$ if $\alpha$ is positive while negative $\alpha$ results in the nonordered state $|\mathbf{v}| = 0$. The coefficients in these equations can be obtained from microscopic models \cite{Bertin2006,Peshkov2014,Farrell2012,Solon2013,Suzuki2015,Bricard2013}. To simplify the problem, we ignore the terms including $\lambda_{2,3}$ and also the diffusive terms that contain the second-order derivative. This condition can be met in a variety of active systems \cite{Souslov2017,Shankar2017,Farrell2012}. 
Effects of $\lambda_{2,3}$ terms can be taken into account by renormalizing $\lambda$ if necessary \cite{Toner1995}. We also assume that the pressure $P$ is proportional to $\rho$ as appropriate for an ideal gas.

Linearizing the Toner-Tu equations around a steady-state solution, we derive an eigenvalue equation for the fluctuations of density and velocity fields, $\delta\rho({\bf r},t) = \rho({\bf r},t) - \rho_{\rm ss}({\bf r})$ and $\delta\mathbf{v}({\bf r},t)=\mathbf{v}({\bf r},t)-\mathbf{v}_{\rm ss}({\bf r})$, respectively, where $\rho_{\rm ss}$ and $\mathbf{v}_{\rm ss}$ represent their steady-state values. We also assume that the steady-state speed $|\mathbf{v}_{\rm ss}|$ is much smaller than the sound velocity $c=\sqrt{P/\rho}$. To clarify the argument, we define the following dimensionless variables: $\mathbf{r}'=\mathbf{r}/a$, $t'=ct/a$, $\delta \rho'({\bf r}',t') = \delta \rho(a{\bf r}',at'/c) / \rho_{\rm ss}(a{\bf r}')$, $\delta \mathbf{v}'({\bf r}',t') = \delta \mathbf{v}(a{\bf r}',at'/c) / c$ and $\mathbf{v}'_{\rm ss}({\bf r}') = \mathbf{v}_{\rm ss}(a{\bf r}') / c$, where $a$ is a characteristic length of a system that we specify as a lattice constant later. 
For the sake of notational simplicity, hereafter we express the dimensionless variables $\mathbf{r}'$, $t'$, $\mathbf{v}'_{\rm ss}$ as $\mathbf{r}$, $t$, $\mathbf{v}_{\rm ss}$. The resulting linearized equation in the frequency domain is
\begin{equation}
\mathcal{H} \left(
  \begin{array}{c}
   \delta \tilde{\rho} \\
   \delta \tilde{v}_x \\
   \delta \tilde{v}_y
  \end{array}
  \right) = \omega
  \left(
  \begin{array}{c}
   \delta \tilde{\rho} \\
   \delta \tilde{v}_x \\
   \delta \tilde{v}_y
  \end{array}
 \right)
 \label{linear-toner-tu-eq}
\end{equation}
with $\cal H$ being the effective Hamiltonian defined as
\begin{equation}
  \mathcal{H} = \left(
  \begin{array}{ccc}
   -i\mathbf{v}_{\rm ss} \cdot \nabla & -i \partial_x & -i \partial_y \\
   -i \partial_x & -i\lambda\mathbf{v}_{\rm ss} \cdot \nabla & 0 \\
   -i \partial_y & 0 & -i\lambda\mathbf{v}_{\rm ss} \cdot \nabla
  \end{array}
  \right),
\end{equation}
where $\delta\tilde{\rho}({\bf r},\omega)$, $\delta \tilde{v}_{x,y}({\bf r},\omega)$ are the Fourier components in the frequency domain. We here omit the spatial variation of $\rho_{\rm ss}$ and the divergence of $\mathbf{v}_{\rm ss}$ as justified in the  Supplemental Materials. We note that, while the coefficient matrix $\mathcal{H}$ can be regarded as the effective Hamiltonian, it can in general be non-Hermitian when the diffusive terms in Eq.~\eqref{toner-tu-eq2} are included. 

\begin{figure}[t]
  \includegraphics[width=70mm, bb=0 000 600 780,clip]{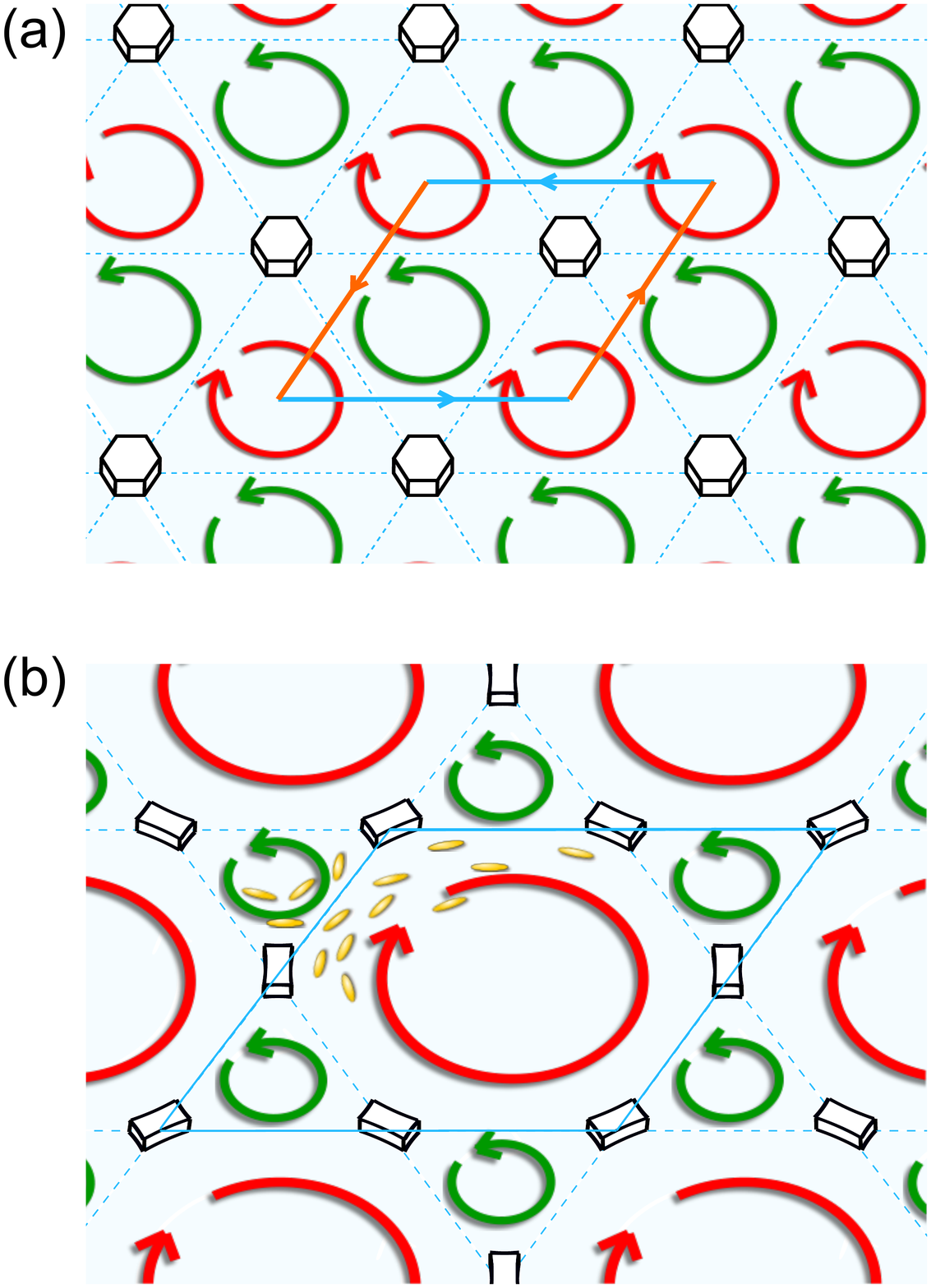}
\caption{\label{fig:lattice} Active particles on the two-dimensional flat space move under the influence of periodically aligned pillars. The red and green curved arrows represent the steady-state flows.
(a) Trivial system with a triangular-lattice geometry.  The line integral along each boundary  of the unit cell (blue and red solid lines) cancels each other because of the periodicity, leading to the vanishing net vorticity. (b) The proposed setup for topological active matter with a kagome-lattice geometry. Blue solid lines indicate the boundary of the unit cell. We set the side length of the unit cell as $a=1$. }
\end{figure}

{\it Absence of net vorticity in active matter.---}
The linearized equation~(\ref{linear-toner-tu-eq}) can be deformed into the Schr\"{o}dinger-like equation \cite{Souslov2017}
\begin{equation}
(-i\nabla-\mathbf{V}_{\rm ss})^2 \delta \tilde{\rho}=\omega^2 \delta \tilde{\rho},
\label{eq:sch}
\end{equation}
where $\mathbf{V}_{\rm ss} = \omega (\lambda+1)\mathbf{v}_{\rm ss}/2$ (see the   Supplemental Materials for the derivation). Equation~\eqref{eq:sch} demonstrates that  $\mathbf{V}_{\rm ss}$ acts as the effective vector potential and its vorticity $\nabla\times{\bf V}_{\rm ss}$ can be interpreted as the effective magnetic field.

We consider active particles without internal degrees of freedom, which reside on a two-dimensional plane with a periodic structure of a unit cell $\Omega$. To avoid  intricate structures, we assume that non-negligibly large defects are absent, i.e., the length of the perimeter of a defect in each unit cell can be neglected with respect to that of the unit-cell boundary $\partial \Omega$. We note that this condition does not preclude possibilities of minuscule defects created by, e.g., thin rods as realized in Ref.~\cite{Nishiguchi2018} or the presence of inhomogeneous potentials relevant to chemotactic bacteria subject to a nonuniform concentration of chemical compounds \cite{Saragosti2011}. 

The net vorticity is then obtained by the integration over the unit cell and can be expressed via the Stokes' theorem as
\begin{equation}
\int_{\Omega} (\nabla \times \mathbf{V}_{\rm ss}) \cdot d \mathbf{S} = \oint_{\partial \Omega} \mathbf{V}_{\rm ss} \cdot d\mathbf{r}.
\label{eq:stokes}
\end{equation}
Due to a periodic structure of the steady flow along the unit-cell boundary $\partial \Omega$, one can show that the line integration in the right-hand side of Eq.~\eqref{eq:stokes} adds up to zero (see Fig.~\ref{fig:lattice} for typical examples), resulting in the vanishing net vorticity. Thus, in the active hydrodynamics of interest here, it is prohibited to realize an analog of the quantum Hall effect, which requires an external magnetic field indicating nonzero net vorticity.  Under the above conditions, we naturally arrive at the conclusion that {\it a topological active matter must be realized as a counterpart of the QAHE}, where the need for the external magnetic field can be mitigated.

We mention the reasons why the counterparts of the quantum Hall effect can be constructed in the previous setups \cite{Dasbiswas2017,Souslov2017,Shankar2017,Souslov2018} despite the above argument. The active particles in Refs.~\cite{Dasbiswas2017,Souslov2018} exhibit self-rotations and thus their internal angular momenta violate our assumption on the absence of internal degrees of freedom. The model in Ref.~\cite{Souslov2018} includes external rotational force, where the Coriolis force acts as an effective Lorentz force. In Ref.~\cite{Souslov2017}, the model includes large defects around which  additional line integrals contribute to extra vorticity in Eq.~\eqref{eq:stokes}.  The curved space is discussed in Ref.~\cite{Shankar2017}; it violates our assumption on flatness of the space.
Altogether, the above effects lead to the emergence of the net effective magnetic field and thus permit realizing the counterparts of the conventional quantum Hall effect.

{\it The minimal model of topological active matter.---} To complete the analogy between the active matter and the QAHE, we propose the minimal model illustrated in Fig.~\ref{fig:lattice}(b).  There, active particles obeying the Toner-Tu equations move under the influence of small pillars located at each site of a kagome lattice \footnote{We assume that, while these pillars can influence the flow of particles, their sizes are small enough such that the conditions discussed above Eq.~\eqref{eq:stokes} are satisfied.}. 
As illustrated in Fig.~\ref{fig:lattice}(b), the steady-state velocity field ${\bf v}_{\rm ss}({\bf r})$ aligns on each boundary of triangular and hexagonal subcells separated by blue solid and dashed lines. Thus, particles in the triangular subcells circulate in the counterclockwise direction (green  arrows) that is opposite to the direction of the particle circulation in the hexagonal subcells (red  arrows) \footnote{The reversed flow $-\mathbf{v}_{\rm ss}$ is also allowed as a steady-state solution;  the initial condition determines which flow can be realized in a system.}, resulting in the vanishing net vorticity. We confirm the emergence of this steady-state flow by performing the particle-based numerical simulation (see Supplementary Materials and Supplementary Movie published with the manuscript).

It is noteworthy that such an ``anticorrelated" velocity profile has been observed in bacterial experiments \cite{Wioland2016,Nishiguchi2018} and also in numerical simulations \cite{Souslov2017,Pearce2015}. In the cases of triangular- and square-lattice structures (cf. Ref.~\cite{Nishiguchi2018} and Fig.~\ref{fig:lattice}(a)), however, only topologically trivial bands can appear due to the absence of a sublattice structure; this is why the kagome-lattice structure (as considered here) is crucial for realizing topological active matter.
To gain physical insights, we point out the analogy between the present system and the QAHE  \cite{Haldane1988,Ohgushi2000}; the collective density fluctuations are influenced by the local vorticity in such a way that electrons propagating through the crystal feel the Berry phase. One can thus expect that  sound modes can exhibit topologically nontrivial bands, as the electronic bands feature the QAHE \cite{Ohgushi2000}. 

\begin{figure}[t]
  \includegraphics[width=70mm, bb=0 0 600 830,clip]{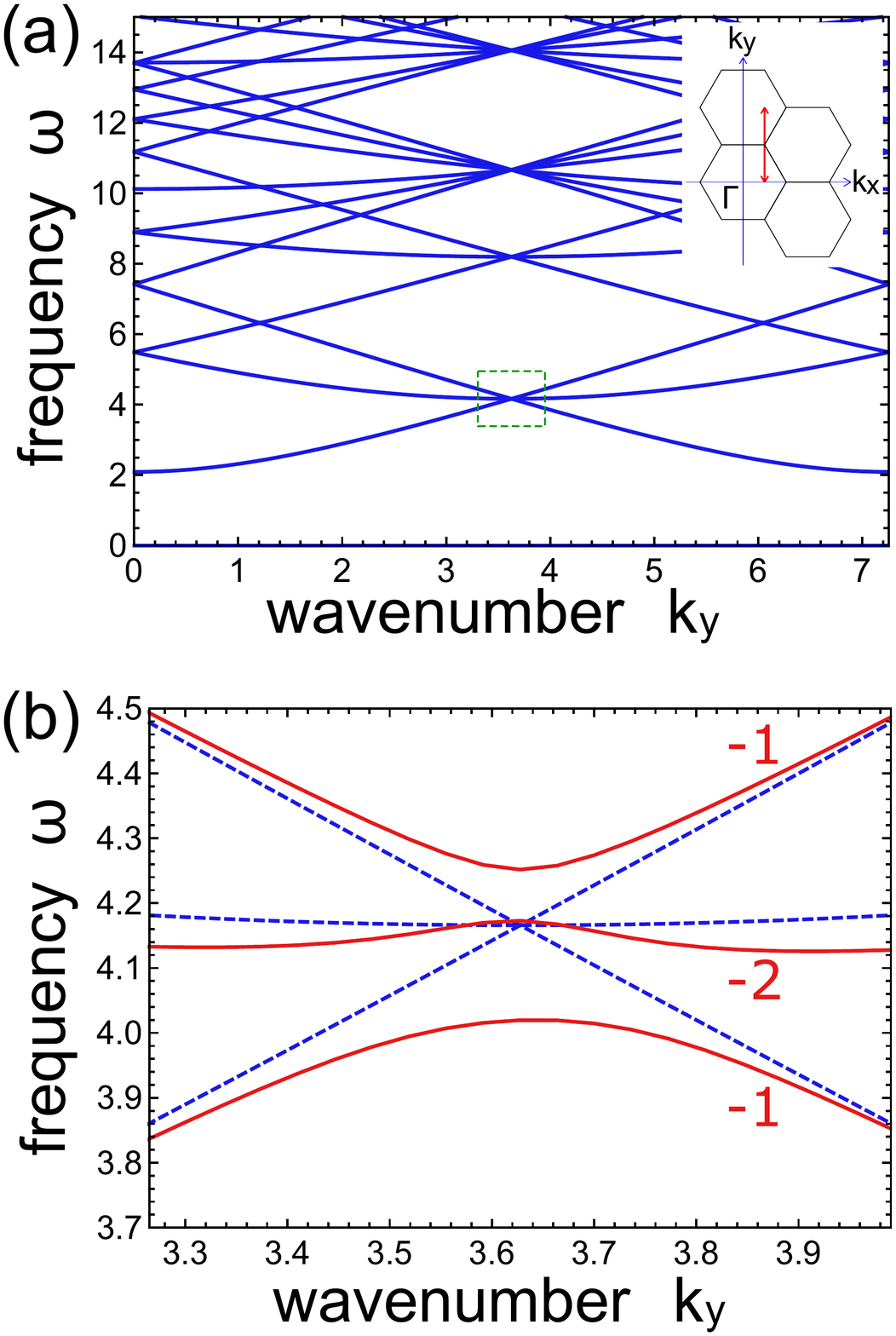}
\caption{\label{fig:2dband} (a) The band structure of the nonordered system, i.e., ${\bf v}_{\rm ss}=0$. The dispersion is plotted along the line at $k_x=2\pi/3$ with varying $k_y$ as indicated by the red arrow in the inset. We impose the twisted boundary conditions, $\psi(\mathbf{x}+\mathbf{a}) = e^{i\mathbf{k}\cdot \mathbf{a}} \psi(\mathbf{x}),$ where $\psi$ is an eigenfunction and $\mathbf{a}$ is a lattice vector. The parameters used are $c=1$, $\rho_{\rm ss}=1$ and $\lambda=0.8$. 
(b) The enlarged view of the band structure in the green dashed box in (a). The orange solid (blue dashed) curves show the results with (without) the steady-state flows. The integer number at each band represents the Chern number.}
\end{figure}

{\it Topological band structure.---} We obtain the bulk dispersion by numerically diagonalizing the effective Hamiltonian $\mathcal{H}$. To obtain accurate results, we add the redundant degree of freedom without affecting the band structure by transforming $\mathcal{H}$ via a unitary matrix (see the Supplementary Materials). This transformation allows for calculations in the basis reflecting the centrosymmetry of the present system. If we calculate the bulk band structure without this redundant degree of freedom, we obtain unphysical $k_y$-independent bands. While this prescription generates redundant eigenstates with eigenvalues of $0$, it does not change any physical properties including the topological feature. We thus analyze the eigenequation
\begin{equation}
\mathcal{H}' \left(
  \begin{array}{c}
   \delta \tilde{\rho} \\
   \delta \tilde{v}_1\\
   \delta \tilde{v}_2 \\
   \delta \tilde{v}_3
  \end{array}
  \right) = \omega
  \left(
  \begin{array}{c}
   \delta \tilde{\rho} \\
   \delta \tilde{v}_1 \\
   \delta \tilde{v}_2 \\
   \delta \tilde{v}_3
  \end{array}
 \right)
\end{equation}
with $\cal H'$ being the effective Hamiltonian in the transformed frame
\eqn{
  \mathcal{H}' \!\!=\!\! -i \left(
  \begin{array}{cccc}
   \mathbf{v}_{\rm ss} \cdot \nabla & \frac{2}{\sqrt{6}} \partial_1 & \frac{2}{\sqrt{6}} \partial_2 & \frac{2}{\sqrt{6}} \partial_3 \\
   \frac{2}{\sqrt{6}} \partial_1 & \frac{2\lambda}{3}\mathbf{v}_{\rm ss} \cdot \nabla & -\frac{\lambda}{3}\mathbf{v}_{\rm ss} \cdot \nabla & -\frac{\lambda}{3}\mathbf{v}_{\rm ss} \cdot \nabla \\
   \frac{2}{\sqrt{6}} \partial_2 & -\frac{\lambda}{3}\mathbf{v}_{\rm ss} \cdot \nabla & \frac{2\lambda}{3}\mathbf{v}_{\rm ss} \cdot \nabla & -\frac{\lambda}{3}\mathbf{v}_{\rm ss} \cdot \nabla \\
   \frac{2}{\sqrt{6}} \partial_3 & -\frac{\lambda}{3}\mathbf{v}_{\rm ss} \cdot \nabla & -\frac{\lambda}{3}\mathbf{v}_{\rm ss} \cdot \nabla & \frac{2\lambda}{3}\mathbf{v}_{\rm ss} \cdot \nabla
  \end{array}
  \right),\nonumber\\
}
where we define the variables as $ \delta \tilde{v}_1\!=\!2\delta \tilde{v}_x/\sqrt{6}\! +\! \delta \tilde{v}_r/\sqrt{3},  $
 $ \delta \tilde{v}_2\!=\! -\delta \tilde{v}_x/\sqrt{6}\!+\!\delta \tilde{v}_y/\sqrt{2}\!+\!\delta \tilde{v}_r/\sqrt{3} $ and 
 $ \delta \tilde{v}_3\!=\! -\delta \tilde{v}_x/\sqrt{6}\!-\!\delta \tilde{v}_y/\sqrt{2}\!+\!\delta \tilde{v}_r/\sqrt{3}.$
 Here, $\delta\tilde{v}_r$ is the redundant degree of freedom. 
 The derivatives denote  
 $ 
 \partial_1\!=\!\partial_x,$ 
$ \partial_2\!=\! -\partial_x/2+{\sqrt{3}}\partial_y/2$ and  
$ \partial_3\!=\! -\partial_x/2-{\sqrt{3}}\partial_y/2$, which correspond to the directions along the grid lines of the kagome lattice (cf. the blue dashed and solid lines in Fig.~\ref{fig:lattice}(b)). Figure~\ref{fig:2dband} shows the band structure of the effective Hamiltonian $\mathcal{H}'$ calculated by the difference method \cite{Smith1985}.  Since the effective Hamiltonian satisfies the particle-hole symmetry, there is a counterpart for each eigenvector whose eigenenergy has the same absolute value and the opposite sign. For the nonordered case $|\mathbf{v}_{\rm ss}|=0$, there are degeneracies at the edges of the first Brillouin zone. Nonzero $|\mathbf{v}_{\rm ss}|$ lifts those degeneracies and opens band gaps, characteristics of topological materials \cite{Haldane1988,Ohgushi2000,Yang2015,Souslov2017,Fleury2016,Rechtsman2013,Khanikaev2013,Shankar2017,Nash2015,Haldane2008}. We note that our effective Hamiltonian does not contain nonderivative terms that are necessary for the original proposal of the quantum anomalous Hall effect \cite{Haldane1988}. The kagome-lattice structure mitigates this requirement \cite{Ohgushi2000}  as it can realize the local flux without next-nearest hoppings.

\begin{figure}[t]
  \includegraphics[width=70mm, bb=0 0 870 580,clip]{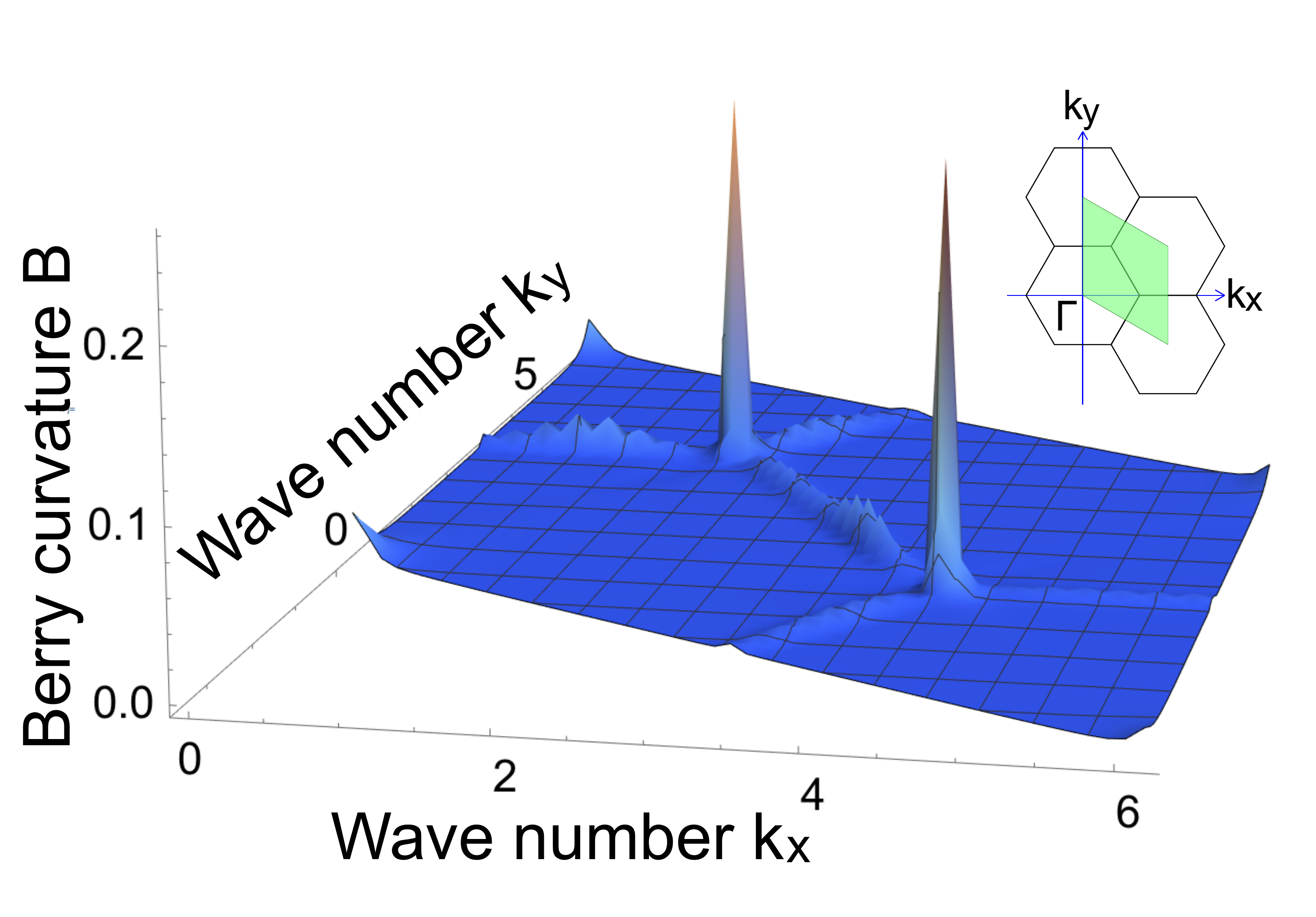}
\caption{\label{fig:berrycur} Berry curvature of a topologically nontrivial band (the middle band with the Chern number $C=-2$ in Fig.~\ref{fig:2dband}). For the sake of visibility, the Berry curvature is plotted by inverting its sign. The parameters used are the same as in Fig.~\ref{fig:2dband}.}
\end{figure}

We confirm that the proposed model exhibits a topologically nontrivial band by calculating the bulk topological invariant, i.e., the Chern number \cite{Simon1983}. Specifically, the Chern number of the \textit{n}-th band is defined as
\begin{equation}
C_n = \frac{1}{2\pi} \int _{\rm BZ} \mathbf{B}_n (\mathbf{q}) \cdot d\mathbf{S},
\end{equation}
where $\mathbf{B}_n (\mathbf{q})= \nabla_{\mathbf{q}} \times \mathbf{A}_n (\mathbf{q})$ is the Berry curvature with $\mathbf{A}_n (\mathbf{q}) = i\mathbf{u}_n(\mathbf{q}) \cdot (\nabla_{\mathbf{q}} \mathbf{u}_n(\mathbf{q}))$ being the Berry connection and $\mathbf{u}_n(\mathbf{q})$ being the \textit{n}-th eigenvector at wavenumber $\mathbf{q}$. We calculate the Berry curvature and the Chern number for each band following the numerical method proposed in Ref.~\cite{Fukui2005}. The calculation shows that many of the bands have nonzero Chern numbers (see e.g., Fig.~\ref{fig:2dband}(b)). Figure~\ref{fig:berrycur} shows the Berry curvature of the topologically nontrivial acoustic band (cf. the middle band with $C=-2$ in Fig.~\ref{fig:2dband}(b)), which exhibits sharp peaks at the edges of the first Brillouin zone. 

\begin{figure}[t]
  \includegraphics[width=70mm, bb=0 0 490 390,clip]{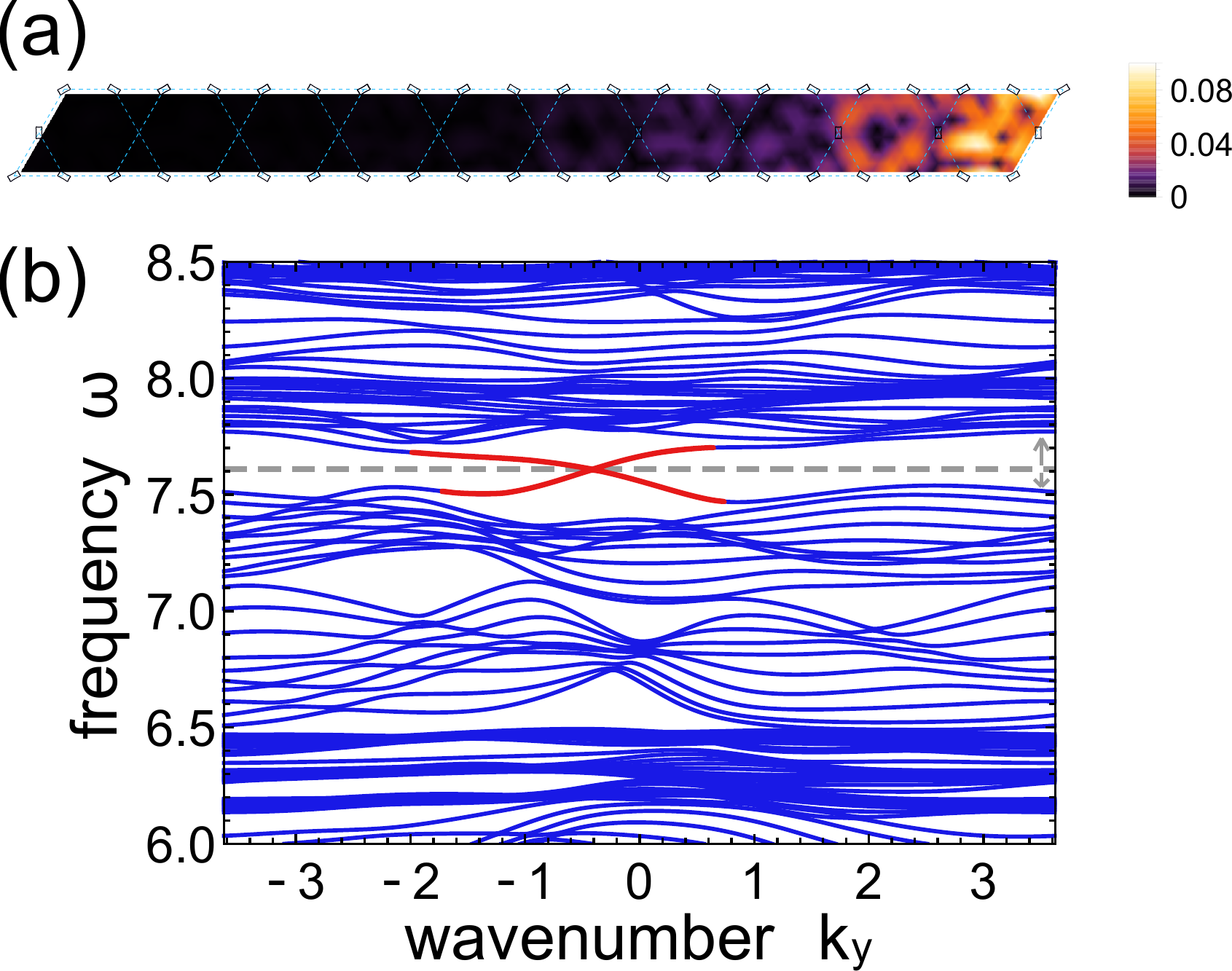}
\caption{\label{fig:edge} (a) Spatial profile of the magnitude of the density fluctuation in an edge mode. We align 10 unit cells with open boundary conditions in the $x$-direction. Periodic boundary conditions are imposed in the $y$-direction. The wavenumber and the frequency are set to be $k_y = -(4\sqrt{3}/15)\pi$ and $\omega = 7.6$, respectively.  (b) The corresponding band structure. The red curves show the dispersion of the edge mode in (a). The gray dashed line indicates the presence of the bulk band gap. }
\end{figure}

The bulk-edge correspondence predicts that the nonzero Chern number accompanies a unidirectional edge mode under  open boundary conditions. While the correspondence has been well established in tight-binding lattice models, it has been recently argued to hold also in the continuum space \cite{Silveirinha2019}. To test the existence of edge modes, we calculate the sound modes for a supercell structure; many identical unit cells are aligned with open boundary conditions in the $x$-direction while the periodic boundary conditions are imposed in the $y$-direction. Figure~\ref{fig:edge} shows the band structure in this setup and the real-space profile of the sound mode at the gap between the topologically distinct bands. The density fluctuation rapidly decreases as we depart from the right end, indicating the presence of the edge mode (Fig.~\ref{fig:edge}a). The edge band connects the lower and upper bulk bands (Fig.~\ref{fig:edge}b). There is one edge mode at the bulk gap as consistent with the sum of the Chern numbers of the bands below the energy gap, $\sum_{n_<} C_{n_<}=-1$.

{\it Summary and Discussions.---} 
We showed that topologically nontrivial bands can arise in active systems without implementing intricate structures, which have been  considered as prerequisites for realizing topological sound modes. 
Due to the vanishing net vorticity of steady-state flows, we pointed out that the quantum anomalous Hall effect provides a natural pathway to realize topological active materials. These findings are supported by numerical calculations of the band structure of the simple model, which is inspired by the flat-band ferromagnet in solid-state systems.

The present study opens several research directions.
Firstly, our results expand possibilities for experimental realizations of topological active systems. Recent experimental developments have enabled one to measure and manipulate polar active matter by using, for example, bacteria. In particular, the experimental setup realized by Ref.~\cite{Nishiguchi2018} is directly relevant to our model except for the lattice structure and thus, our theoretical results can be tested with current experimental techniques. 
Secondly, our work suggests a simple and general way to construct topological active matter by designing its periodic structure (steady-state flow) with making the analogy to a profile of a tight-binding lattice (gauge field) relevant to electronic topological materials. 

Thirdly, besides technological applications, one major motivation in the field of active matter is to advance our understanding of emergent nonequilibrium phenomena in biological systems.
Biological systems modeled as active matter include, for example, cells, molecular motors, and cytoskeletons \cite{Marchetti2013,Vicsek2012}. Topological edge modes may play an important role in various biological functionalities, which are often robust to disorder. 

Finally, while we neglect the diffusive terms in the Toner-Tu equation~\eqref{toner-tu-eq2}, they can in general make the effective Hamiltonian non-Hermitian and suppress the high-wavenumber modes. It is worthwhile to explore non-Hermitian topological phenomena in active systems; of particular interest is an exotic topological feature that has no counterpart in Hermitian systems \cite{Lee2016,Leykam2017,Kunst2018,Yao2018,Gong2018}. In particular,  asymmetrical flows in active matter may allow one to realize the non-Hermitian skin effect \cite{Kunst2018,Yao2018,Longhi2019} and the quasiedge modes \cite{Gong2018}. Such a feature could lead to an emergence of novel functionalities unique to active matter. 
We hope that our work stimulates further studies in these directions.

We thank Shunsuke Furukawa, Ryusuke Hamazaki, Takahiro Sagawa, Masahito Ueda, Daiki Nishiguchi, and Kazumasa Takeuchi for useful discussions. Y.A. acknowledges support from the Japan Society for the Promotion of Science through Program for Leading Graduate Schools (ALPS) and Grant No.~JP16J03613 and JP19K23424.

\bibliography{reference}

\begin{thebibliography}{67}%
\makeatletter
\providecommand \@ifxundefined [1]{%
 \@ifx{#1\undefined}
}%
\providecommand \@ifnum [1]{%
 \ifnum #1\expandafter \@firstoftwo
 \else \expandafter \@secondoftwo
 \fi
}%
\providecommand \@ifx [1]{%
 \ifx #1\expandafter \@firstoftwo
 \else \expandafter \@secondoftwo
 \fi
}%
\providecommand \natexlab [1]{#1}%
\providecommand \enquote  [1]{``#1''}%
\providecommand \bibnamefont  [1]{#1}%
\providecommand \bibfnamefont [1]{#1}%
\providecommand \citenamefont [1]{#1}%
\providecommand \href@noop [0]{\@secondoftwo}%
\providecommand \href [0]{\begingroup \@sanitize@url \@href}%
\providecommand \@href[1]{\@@startlink{#1}\@@href}%
\providecommand \@@href[1]{\endgroup#1\@@endlink}%
\providecommand \@sanitize@url [0]{\catcode `\\12\catcode `\$12\catcode
  `\&12\catcode `\#12\catcode `\^12\catcode `\_12\catcode `\%12\relax}%
\providecommand \@@startlink[1]{}%
\providecommand \@@endlink[0]{}%
\providecommand \url  [0]{\begingroup\@sanitize@url \@url }%
\providecommand \@url [1]{\endgroup\@href {#1}{\urlprefix }}%
\providecommand \urlprefix  [0]{URL }%
\providecommand \Eprint [0]{\href }%
\providecommand \doibase [0]{http://dx.doi.org/}%
\providecommand \selectlanguage [0]{\@gobble}%
\providecommand \bibinfo  [0]{\@secondoftwo}%
\providecommand \bibfield  [0]{\@secondoftwo}%
\providecommand \translation [1]{[#1]}%
\providecommand \BibitemOpen [0]{}%
\providecommand \bibitemStop [0]{}%
\providecommand \bibitemNoStop [0]{.\EOS\space}%
\providecommand \EOS [0]{\spacefactor3000\relax}%
\providecommand \BibitemShut  [1]{\csname bibitem#1\endcsname}%
\let\auto@bib@innerbib\@empty
\bibitem [{\citenamefont {Thouless}\ \emph {et~al.}(1982)\citenamefont
  {Thouless}, \citenamefont {Kohmoto}, \citenamefont {Nightingale},\ and\
  \citenamefont {Nijs}}]{Thouless1982}%
  \BibitemOpen
  \bibfield  {author} {\bibinfo {author} {\bibfnamefont {D.~J.}\ \bibnamefont
  {Thouless}}, \bibinfo {author} {\bibfnamefont {M.}~\bibnamefont {Kohmoto}},
  \bibinfo {author} {\bibfnamefont {M.~P.}\ \bibnamefont {Nightingale}}, \ and\
  \bibinfo {author} {\bibfnamefont {M.~D.}\ \bibnamefont {Nijs}},\ }\href
  {\doibase 10.1103/PhysRevLett.49.405} {\bibfield  {journal} {\bibinfo
  {journal} {Phys. Rev. Lett.}\ }\textbf {\bibinfo {volume} {49}},\ \bibinfo
  {pages} {405} (\bibinfo {year} {1982})}\BibitemShut {NoStop}%
\bibitem [{\citenamefont {Simon}(1983)}]{Simon1983}%
  \BibitemOpen
  \bibfield  {author} {\bibinfo {author} {\bibfnamefont {B.}~\bibnamefont
  {Simon}},\ }\href {\doibase 10.1103/PhysRevLett.51.2167} {\bibfield
  {journal} {\bibinfo  {journal} {Phys. Rev. Lett.}\ }\textbf {\bibinfo
  {volume} {51}},\ \bibinfo {pages} {2167} (\bibinfo {year}
  {1983})}\BibitemShut {NoStop}%
\bibitem [{\citenamefont {{F. D. M. Haldane}}(1988)}]{Haldane1988}%
  \BibitemOpen
  \bibfield  {author} {\bibinfo {author} {\bibnamefont {{F. D. M. Haldane}}},\
  }\href {\doibase 10.1103/PhysRevLett.61.2015} {\bibfield  {journal} {\bibinfo
   {journal} {Phys. Rev. Lett.}\ }\textbf {\bibinfo {volume} {61}},\ \bibinfo
  {pages} {2015} (\bibinfo {year} {1988})}\BibitemShut {NoStop}%
\bibitem [{\citenamefont {Kane}\ and\ \citenamefont {Mele}(2005)}]{Kane2005}%
  \BibitemOpen
  \bibfield  {author} {\bibinfo {author} {\bibfnamefont {C.~L.}\ \bibnamefont
  {Kane}}\ and\ \bibinfo {author} {\bibfnamefont {E.~J.}\ \bibnamefont
  {Mele}},\ }\href {\doibase 10.1103/PhysRevLett.95.226801} {\bibfield
  {journal} {\bibinfo  {journal} {Phys. Rev. Lett.}\ }\textbf {\bibinfo
  {volume} {95}},\ \bibinfo {pages} {226801} (\bibinfo {year}
  {2005})}\BibitemShut {NoStop}%
\bibitem [{\citenamefont {Hasan}\ and\ \citenamefont {Kane}(2010)}]{Hasan2010}%
  \BibitemOpen
  \bibfield  {author} {\bibinfo {author} {\bibfnamefont {M.~Z.}\ \bibnamefont
  {Hasan}}\ and\ \bibinfo {author} {\bibfnamefont {C.~L.}\ \bibnamefont
  {Kane}},\ }\href {\doibase 10.1103/RevModPhys.82.3045} {\bibfield  {journal}
  {\bibinfo  {journal} {Rev. Mod. Phys.}\ }\textbf {\bibinfo {volume} {82}},\
  \bibinfo {pages} {3045} (\bibinfo {year} {2010})}\BibitemShut {NoStop}%
\bibitem [{\citenamefont {Qi}\ and\ \citenamefont {Zhang}(2011)}]{Qi2011}%
  \BibitemOpen
  \bibfield  {author} {\bibinfo {author} {\bibfnamefont {X.~L.}\ \bibnamefont
  {Qi}}\ and\ \bibinfo {author} {\bibfnamefont {S.~C.}\ \bibnamefont {Zhang}},\
  }\href {\doibase 10.1103/RevModPhys.83.1057} {\bibfield  {journal} {\bibinfo
  {journal} {Rev. Mod. Phys.}\ }\textbf {\bibinfo {volume} {83}},\ \bibinfo
  {pages} {1057} (\bibinfo {year} {2011})}\BibitemShut {NoStop}%
\bibitem [{\citenamefont {Haldane}\ and\ \citenamefont
  {Raghu}(2008)}]{Haldane2008}%
  \BibitemOpen
  \bibfield  {author} {\bibinfo {author} {\bibfnamefont {F.~D.~M.}\
  \bibnamefont {Haldane}}\ and\ \bibinfo {author} {\bibfnamefont
  {S.}~\bibnamefont {Raghu}},\ }\href {\doibase 10.1103/PhysRevLett.100.013904}
  {\bibfield  {journal} {\bibinfo  {journal} {Phys. Rev. Lett.}\ }\textbf
  {\bibinfo {volume} {100}},\ \bibinfo {pages} {013904} (\bibinfo {year}
  {2008})}\BibitemShut {NoStop}%
\bibitem [{\citenamefont {Lu}\ \emph {et~al.}(2014)\citenamefont {Lu},
  \citenamefont {Joannopoulos},\ and\ \citenamefont
  {Solja{\v{c}}i{\'{c}}}}]{Lu2014}%
  \BibitemOpen
  \bibfield  {author} {\bibinfo {author} {\bibfnamefont {L.}~\bibnamefont
  {Lu}}, \bibinfo {author} {\bibfnamefont {J.~D.}\ \bibnamefont
  {Joannopoulos}}, \ and\ \bibinfo {author} {\bibfnamefont {M.}~\bibnamefont
  {Solja{\v{c}}i{\'{c}}}},\ }\href {\doibase 10.1038/nphoton.2014.248}
  {\bibfield  {journal} {\bibinfo  {journal} {Nat. Photonics}\ }\textbf
  {\bibinfo {volume} {8}},\ \bibinfo {pages} {821} (\bibinfo {year}
  {2014})}\BibitemShut {NoStop}%
\bibitem [{\citenamefont {Rechtsman}\ \emph {et~al.}(2013)\citenamefont
  {Rechtsman}, \citenamefont {Zeuner}, \citenamefont {Plotnik}, \citenamefont
  {Lumer}, \citenamefont {Podolsky}, \citenamefont {Dreisow}, \citenamefont
  {Nolte}, \citenamefont {Segev},\ and\ \citenamefont
  {Szameit}}]{Rechtsman2013}%
  \BibitemOpen
  \bibfield  {author} {\bibinfo {author} {\bibfnamefont {M.~C.}\ \bibnamefont
  {Rechtsman}}, \bibinfo {author} {\bibfnamefont {J.~M.}\ \bibnamefont
  {Zeuner}}, \bibinfo {author} {\bibfnamefont {Y.}~\bibnamefont {Plotnik}},
  \bibinfo {author} {\bibfnamefont {Y.}~\bibnamefont {Lumer}}, \bibinfo
  {author} {\bibfnamefont {D.}~\bibnamefont {Podolsky}}, \bibinfo {author}
  {\bibfnamefont {F.}~\bibnamefont {Dreisow}}, \bibinfo {author} {\bibfnamefont
  {S.}~\bibnamefont {Nolte}}, \bibinfo {author} {\bibfnamefont
  {M.}~\bibnamefont {Segev}}, \ and\ \bibinfo {author} {\bibfnamefont
  {A.}~\bibnamefont {Szameit}},\ }\href {\doibase 10.1038/nature12066}
  {\bibfield  {journal} {\bibinfo  {journal} {Nature}\ }\textbf {\bibinfo
  {volume} {496}},\ \bibinfo {pages} {196} (\bibinfo {year}
  {2013})}\BibitemShut {NoStop}%
\bibitem [{\citenamefont {Khanikaev}\ \emph {et~al.}(2013)\citenamefont
  {Khanikaev}, \citenamefont {{Hossein Mousavi}}, \citenamefont {Tse},
  \citenamefont {Kargarian}, \citenamefont {MacDonald},\ and\ \citenamefont
  {Shvets}}]{Khanikaev2013}%
  \BibitemOpen
  \bibfield  {author} {\bibinfo {author} {\bibfnamefont {A.~B.}\ \bibnamefont
  {Khanikaev}}, \bibinfo {author} {\bibfnamefont {S.}~\bibnamefont {{Hossein
  Mousavi}}}, \bibinfo {author} {\bibfnamefont {W.~K.}\ \bibnamefont {Tse}},
  \bibinfo {author} {\bibfnamefont {M.}~\bibnamefont {Kargarian}}, \bibinfo
  {author} {\bibfnamefont {A.~H.}\ \bibnamefont {MacDonald}}, \ and\ \bibinfo
  {author} {\bibfnamefont {G.}~\bibnamefont {Shvets}},\ }\href {\doibase
  10.1038/nmat3520} {\bibfield  {journal} {\bibinfo  {journal} {Nat. Mater.}\
  }\textbf {\bibinfo {volume} {12}},\ \bibinfo {pages} {233} (\bibinfo {year}
  {2013})}\BibitemShut {NoStop}%
\bibitem [{\citenamefont {Huber}(2016)}]{Huber2016}%
  \BibitemOpen
  \bibfield  {author} {\bibinfo {author} {\bibfnamefont {S.~D.}\ \bibnamefont
  {Huber}},\ }\href {\doibase 10.1038/nphys3801} {\bibfield  {journal}
  {\bibinfo  {journal} {Nat. Phys.}\ }\textbf {\bibinfo {volume} {12}},\
  \bibinfo {pages} {621} (\bibinfo {year} {2016})}\BibitemShut {NoStop}%
\bibitem [{\citenamefont {Nash}\ \emph {et~al.}(2015)\citenamefont {Nash},
  \citenamefont {Kleckner}, \citenamefont {Read}, \citenamefont {Vitelli},
  \citenamefont {Turner},\ and\ \citenamefont {Irvine}}]{Nash2015}%
  \BibitemOpen
  \bibfield  {author} {\bibinfo {author} {\bibfnamefont {L.~M.}\ \bibnamefont
  {Nash}}, \bibinfo {author} {\bibfnamefont {D.}~\bibnamefont {Kleckner}},
  \bibinfo {author} {\bibfnamefont {A.}~\bibnamefont {Read}}, \bibinfo {author}
  {\bibfnamefont {V.}~\bibnamefont {Vitelli}}, \bibinfo {author} {\bibfnamefont
  {A.~M.}\ \bibnamefont {Turner}}, \ and\ \bibinfo {author} {\bibfnamefont
  {W.~T.~M.}\ \bibnamefont {Irvine}},\ }\href {\doibase
  10.1073/pnas.1507413112} {\bibfield  {journal} {\bibinfo  {journal} {Proc.
  Natl. Acad. Sci. U.S.A.}\ }\textbf {\bibinfo {volume} {112}},\ \bibinfo
  {pages} {14495} (\bibinfo {year} {2015})}\BibitemShut {NoStop}%
\bibitem [{\citenamefont {S{\"{u}}sstrunk}\ and\ \citenamefont
  {Huber}(2015)}]{Susstrunk2015}%
  \BibitemOpen
  \bibfield  {author} {\bibinfo {author} {\bibfnamefont {R.}~\bibnamefont
  {S{\"{u}}sstrunk}}\ and\ \bibinfo {author} {\bibfnamefont {S.~D.}\
  \bibnamefont {Huber}},\ }\href {\doibase 10.1126/science.aab0239} {\bibfield
  {journal} {\bibinfo  {journal} {Science}\ }\textbf {\bibinfo {volume}
  {349}},\ \bibinfo {pages} {47} (\bibinfo {year} {2015})}\BibitemShut
  {NoStop}%
\bibitem [{\citenamefont {Kane}\ and\ \citenamefont
  {Lubensky}(2014)}]{Kane2014}%
  \BibitemOpen
  \bibfield  {author} {\bibinfo {author} {\bibfnamefont {C.~L.}\ \bibnamefont
  {Kane}}\ and\ \bibinfo {author} {\bibfnamefont {T.~C.}\ \bibnamefont
  {Lubensky}},\ }\href {\doibase 10.1038/nphys2835} {\bibfield  {journal}
  {\bibinfo  {journal} {Nat. Phys.}\ }\textbf {\bibinfo {volume} {10}},\
  \bibinfo {pages} {39} (\bibinfo {year} {2014})}\BibitemShut {NoStop}%
\bibitem [{\citenamefont {Yang}\ \emph {et~al.}(2015)\citenamefont {Yang},
  \citenamefont {Gao}, \citenamefont {Shi}, \citenamefont {Lin}, \citenamefont
  {Gao}, \citenamefont {Chong},\ and\ \citenamefont {Zhang}}]{Yang2015}%
  \BibitemOpen
  \bibfield  {author} {\bibinfo {author} {\bibfnamefont {Z.}~\bibnamefont
  {Yang}}, \bibinfo {author} {\bibfnamefont {F.}~\bibnamefont {Gao}}, \bibinfo
  {author} {\bibfnamefont {X.}~\bibnamefont {Shi}}, \bibinfo {author}
  {\bibfnamefont {X.}~\bibnamefont {Lin}}, \bibinfo {author} {\bibfnamefont
  {Z.}~\bibnamefont {Gao}}, \bibinfo {author} {\bibfnamefont {Y.}~\bibnamefont
  {Chong}}, \ and\ \bibinfo {author} {\bibfnamefont {B.}~\bibnamefont
  {Zhang}},\ }\href {\doibase 10.1103/PhysRevLett.114.114301} {\bibfield
  {journal} {\bibinfo  {journal} {Phys. Rev. Lett.}\ }\textbf {\bibinfo
  {volume} {114}},\ \bibinfo {pages} {114301} (\bibinfo {year}
  {2015})}\BibitemShut {NoStop}%
\bibitem [{\citenamefont {Fleury}\ \emph {et~al.}(2016)\citenamefont {Fleury},
  \citenamefont {Khanikaev},\ and\ \citenamefont {Al{\`{u}}}}]{Fleury2016}%
  \BibitemOpen
  \bibfield  {author} {\bibinfo {author} {\bibfnamefont {R.}~\bibnamefont
  {Fleury}}, \bibinfo {author} {\bibfnamefont {A.~B.}\ \bibnamefont
  {Khanikaev}}, \ and\ \bibinfo {author} {\bibfnamefont {A.}~\bibnamefont
  {Al{\`{u}}}},\ }\href {\doibase 10.1038/ncomms11744} {\bibfield  {journal}
  {\bibinfo  {journal} {Nat. Commun.}\ }\textbf {\bibinfo {volume} {7}},\
  \bibinfo {pages} {11744} (\bibinfo {year} {2016})}\BibitemShut {NoStop}%
\bibitem [{\citenamefont {Delplace}\ \emph {et~al.}(2017)\citenamefont
  {Delplace}, \citenamefont {Marston},\ and\ \citenamefont
  {Venaille}}]{Delplace2017}%
  \BibitemOpen
  \bibfield  {author} {\bibinfo {author} {\bibfnamefont {P.}~\bibnamefont
  {Delplace}}, \bibinfo {author} {\bibfnamefont {J.~B.}\ \bibnamefont
  {Marston}}, \ and\ \bibinfo {author} {\bibfnamefont {A.}~\bibnamefont
  {Venaille}},\ }\href {\doibase 10.1126/science.aan8819} {\bibfield  {journal}
  {\bibinfo  {journal} {Science}\ }\textbf {\bibinfo {volume} {358}},\ \bibinfo
  {pages} {1075} (\bibinfo {year} {2017})}\BibitemShut {NoStop}%
\bibitem [{\citenamefont {Murugan}\ and\ \citenamefont
  {Vaikuntanathan}(2017)}]{Murugan2017}%
  \BibitemOpen
  \bibfield  {author} {\bibinfo {author} {\bibfnamefont {A.}~\bibnamefont
  {Murugan}}\ and\ \bibinfo {author} {\bibfnamefont {S.}~\bibnamefont
  {Vaikuntanathan}},\ }\href {\doibase 10.1038/ncomms13881} {\bibfield
  {journal} {\bibinfo  {journal} {Nat. Commun.}\ }\textbf {\bibinfo {volume}
  {8}},\ \bibinfo {pages} {13881} (\bibinfo {year} {2017})}\BibitemShut
  {NoStop}%
\bibitem [{\citenamefont {Dasbiswas}\ \emph {et~al.}(2018)\citenamefont
  {Dasbiswas}, \citenamefont {Mandadapu},\ and\ \citenamefont
  {Vaikuntanathan}}]{Dasbiswas2017}%
  \BibitemOpen
  \bibfield  {author} {\bibinfo {author} {\bibfnamefont {K.}~\bibnamefont
  {Dasbiswas}}, \bibinfo {author} {\bibfnamefont {K.~K.}\ \bibnamefont
  {Mandadapu}}, \ and\ \bibinfo {author} {\bibfnamefont {S.}~\bibnamefont
  {Vaikuntanathan}},\ }\href {\doibase 10.1073/pnas.1721096115} {\bibfield
  {journal} {\bibinfo  {journal} {Proc. Natl. Acad. Sci. U.S.A.}\ }\textbf
  {\bibinfo {volume} {115}},\ \bibinfo {pages} {39} (\bibinfo {year}
  {2018})}\BibitemShut {NoStop}%
\bibitem [{\citenamefont {Brugu{\'{e}}s}\ and\ \citenamefont
  {Needleman}(2014)}]{Brugues2014}%
  \BibitemOpen
  \bibfield  {author} {\bibinfo {author} {\bibfnamefont {J.}~\bibnamefont
  {Brugu{\'{e}}s}}\ and\ \bibinfo {author} {\bibfnamefont {D.}~\bibnamefont
  {Needleman}},\ }\href {\doibase 10.1073/pnas.1409404111} {\bibfield
  {journal} {\bibinfo  {journal} {Proc. Natl. Acad. Sci. U.S.A.}\ }\textbf
  {\bibinfo {volume} {111}},\ \bibinfo {pages} {18496} (\bibinfo {year}
  {2014})}\BibitemShut {NoStop}%
\bibitem [{\citenamefont {Saw}\ \emph {et~al.}(2017)\citenamefont {Saw},
  \citenamefont {Doostmohammadi}, \citenamefont {Nier}, \citenamefont
  {Kocgozlu}, \citenamefont {Thampi}, \citenamefont {Toyama}, \citenamefont
  {Marcq}, \citenamefont {Lim}, \citenamefont {Yeomans},\ and\ \citenamefont
  {Ladoux}}]{Saw2017}%
  \BibitemOpen
  \bibfield  {author} {\bibinfo {author} {\bibfnamefont {T.~B.}\ \bibnamefont
  {Saw}}, \bibinfo {author} {\bibfnamefont {A.}~\bibnamefont {Doostmohammadi}},
  \bibinfo {author} {\bibfnamefont {V.}~\bibnamefont {Nier}}, \bibinfo {author}
  {\bibfnamefont {L.}~\bibnamefont {Kocgozlu}}, \bibinfo {author}
  {\bibfnamefont {S.}~\bibnamefont {Thampi}}, \bibinfo {author} {\bibfnamefont
  {Y.}~\bibnamefont {Toyama}}, \bibinfo {author} {\bibfnamefont
  {P.}~\bibnamefont {Marcq}}, \bibinfo {author} {\bibfnamefont {C.~T.}\
  \bibnamefont {Lim}}, \bibinfo {author} {\bibfnamefont {J.~M.}\ \bibnamefont
  {Yeomans}}, \ and\ \bibinfo {author} {\bibfnamefont {B.}~\bibnamefont
  {Ladoux}},\ }\href {\doibase 10.1038/nature21718} {\bibfield  {journal}
  {\bibinfo  {journal} {Nature}\ }\textbf {\bibinfo {volume} {544}},\ \bibinfo
  {pages} {212} (\bibinfo {year} {2017})}\BibitemShut {NoStop}%
\bibitem [{\citenamefont {Kawaguchi}\ \emph {et~al.}(2017)\citenamefont
  {Kawaguchi}, \citenamefont {Kageyama},\ and\ \citenamefont
  {Sano}}]{Kawaguchi2017}%
  \BibitemOpen
  \bibfield  {author} {\bibinfo {author} {\bibfnamefont {K.}~\bibnamefont
  {Kawaguchi}}, \bibinfo {author} {\bibfnamefont {R.}~\bibnamefont {Kageyama}},
  \ and\ \bibinfo {author} {\bibfnamefont {M.}~\bibnamefont {Sano}},\ }\href
  {\doibase 10.1038/nature22321} {\bibfield  {journal} {\bibinfo  {journal}
  {Nature}\ }\textbf {\bibinfo {volume} {545}},\ \bibinfo {pages} {327}
  (\bibinfo {year} {2017})}\BibitemShut {NoStop}%
\bibitem [{\citenamefont {Ganguly}\ and\ \citenamefont
  {Chaudhuri}(2013)}]{Ganguly2013}%
  \BibitemOpen
  \bibfield  {author} {\bibinfo {author} {\bibfnamefont {C.}~\bibnamefont
  {Ganguly}}\ and\ \bibinfo {author} {\bibfnamefont {D.}~\bibnamefont
  {Chaudhuri}},\ }\href {\doibase 10.1103/PhysRevE.88.032102} {\bibfield
  {journal} {\bibinfo  {journal} {Phys. Rev. E}\ }\textbf {\bibinfo {volume}
  {88}},\ \bibinfo {pages} {032102} (\bibinfo {year} {2013})}\BibitemShut
  {NoStop}%
\bibitem [{\citenamefont {Fodor}\ \emph {et~al.}(2016)\citenamefont {Fodor},
  \citenamefont {Nardini}, \citenamefont {Cates}, \citenamefont {Tailleur},
  \citenamefont {Visco},\ and\ \citenamefont {{Van Wijland}}}]{Fodor2016}%
  \BibitemOpen
  \bibfield  {author} {\bibinfo {author} {\bibfnamefont {{\'{E}}.}~\bibnamefont
  {Fodor}}, \bibinfo {author} {\bibfnamefont {C.}~\bibnamefont {Nardini}},
  \bibinfo {author} {\bibfnamefont {M.~E.}\ \bibnamefont {Cates}}, \bibinfo
  {author} {\bibfnamefont {J.}~\bibnamefont {Tailleur}}, \bibinfo {author}
  {\bibfnamefont {P.}~\bibnamefont {Visco}}, \ and\ \bibinfo {author}
  {\bibfnamefont {F.}~\bibnamefont {{Van Wijland}}},\ }\href {\doibase
  10.1103/PhysRevLett.117.038103} {\bibfield  {journal} {\bibinfo  {journal}
  {Phys. Rev. Lett.}\ }\textbf {\bibinfo {volume} {117}},\ \bibinfo {pages}
  {038103} (\bibinfo {year} {2016})}\BibitemShut {NoStop}%
\bibitem [{\citenamefont {Krishnamurthy}\ \emph {et~al.}(2016)\citenamefont
  {Krishnamurthy}, \citenamefont {Ghosh}, \citenamefont {Chatterji},
  \citenamefont {Ganapathy},\ and\ \citenamefont {Sood}}]{Krishnamurthy2016}%
  \BibitemOpen
  \bibfield  {author} {\bibinfo {author} {\bibfnamefont {S.}~\bibnamefont
  {Krishnamurthy}}, \bibinfo {author} {\bibfnamefont {S.}~\bibnamefont
  {Ghosh}}, \bibinfo {author} {\bibfnamefont {D.}~\bibnamefont {Chatterji}},
  \bibinfo {author} {\bibfnamefont {R.}~\bibnamefont {Ganapathy}}, \ and\
  \bibinfo {author} {\bibfnamefont {A.~K.}\ \bibnamefont {Sood}},\ }\href
  {\doibase 10.1038/nphys3870} {\bibfield  {journal} {\bibinfo  {journal} {Nat.
  Phys.}\ }\textbf {\bibinfo {volume} {12}},\ \bibinfo {pages} {1134} (\bibinfo
  {year} {2016})}\BibitemShut {NoStop}%
\bibitem [{\citenamefont {Mandal}\ \emph {et~al.}(2017)\citenamefont {Mandal},
  \citenamefont {Klymko},\ and\ \citenamefont {DeWeese}}]{Mandal2017}%
  \BibitemOpen
  \bibfield  {author} {\bibinfo {author} {\bibfnamefont {D.}~\bibnamefont
  {Mandal}}, \bibinfo {author} {\bibfnamefont {K.}~\bibnamefont {Klymko}}, \
  and\ \bibinfo {author} {\bibfnamefont {M.~R.}\ \bibnamefont {DeWeese}},\
  }\href {\doibase 10.1103/PhysRevLett.119.258001} {\bibfield  {journal}
  {\bibinfo  {journal} {Phys. Rev. Lett.}\ }\textbf {\bibinfo {volume} {119}},\
  \bibinfo {pages} {258001} (\bibinfo {year} {2017})}\BibitemShut {NoStop}%
\bibitem [{\citenamefont {Pietzonka}\ and\ \citenamefont
  {Seifert}()}]{Pietzonka2018}%
  \BibitemOpen
  \bibfield  {author} {\bibinfo {author} {\bibfnamefont {P.}~\bibnamefont
  {Pietzonka}}\ and\ \bibinfo {author} {\bibfnamefont {U.}~\bibnamefont
  {Seifert}},\ }\href {\doibase 10.1088/1751-8121/aa91b9} {\bibinfo  {journal}
  {J. Phys. A}\ ,\ \bibinfo {pages} {01LT01}}\BibitemShut {NoStop}%
\bibitem [{\citenamefont {Martin}\ \emph {et~al.}(2018)\citenamefont {Martin},
  \citenamefont {Nardini}, \citenamefont {Cates},\ and\ \citenamefont
  {Fodor}}]{Martin2018}%
  \BibitemOpen
\bibfield  {journal} {  }\bibfield  {author} {\bibinfo {author} {\bibfnamefont
  {D.}~\bibnamefont {Martin}}, \bibinfo {author} {\bibfnamefont
  {C.}~\bibnamefont {Nardini}}, \bibinfo {author} {\bibfnamefont {M.~E.}\
  \bibnamefont {Cates}}, \ and\ \bibinfo {author} {\bibfnamefont
  {{\'{E}}.}~\bibnamefont {Fodor}},\ }\href {\doibase
  10.1209/0295-5075/121/60005} {\bibfield  {journal} {\bibinfo  {journal}
  {EPL}\ }\textbf {\bibinfo {volume} {121}},\ \bibinfo {pages} {60005}
  (\bibinfo {year} {2018})}\BibitemShut {NoStop}%
\bibitem [{\citenamefont {Vicsek}\ \emph {et~al.}(1995)\citenamefont {Vicsek},
  \citenamefont {Czir{\'o}k}, \citenamefont {Ben-Jacob}, \citenamefont
  {Cohen},\ and\ \citenamefont {Shochet}}]{Vicsek1995}%
  \BibitemOpen
  \bibfield  {author} {\bibinfo {author} {\bibfnamefont {T.}~\bibnamefont
  {Vicsek}}, \bibinfo {author} {\bibfnamefont {A.}~\bibnamefont {Czir{\'o}k}},
  \bibinfo {author} {\bibfnamefont {E.}~\bibnamefont {Ben-Jacob}}, \bibinfo
  {author} {\bibfnamefont {I.}~\bibnamefont {Cohen}}, \ and\ \bibinfo {author}
  {\bibfnamefont {O.}~\bibnamefont {Shochet}},\ }\href {\doibase
  10.1103/PhysRevLett.75.1226} {\bibfield  {journal} {\bibinfo  {journal}
  {Phys. Rev. Lett.}\ }\textbf {\bibinfo {volume} {75}},\ \bibinfo {pages}
  {1226} (\bibinfo {year} {1995})}\BibitemShut {NoStop}%
\bibitem [{\citenamefont {Vicsek}\ and\ \citenamefont
  {Zafeiris}(2012)}]{Vicsek2012}%
  \BibitemOpen
  \bibfield  {author} {\bibinfo {author} {\bibfnamefont {T.}~\bibnamefont
  {Vicsek}}\ and\ \bibinfo {author} {\bibfnamefont {A.}~\bibnamefont
  {Zafeiris}},\ }\href {\doibase 10.1016/j.physrep.2012.03.004} {\bibfield
  {journal} {\bibinfo  {journal} {Phys. Rep.}\ }\textbf {\bibinfo {volume}
  {517}},\ \bibinfo {pages} {71} (\bibinfo {year} {2012})}\BibitemShut
  {NoStop}%
\bibitem [{\citenamefont {Jiang}\ \emph {et~al.}(2010)\citenamefont {Jiang},
  \citenamefont {Yoshinaga},\ and\ \citenamefont {Sano}}]{Jiang2010}%
  \BibitemOpen
  \bibfield  {author} {\bibinfo {author} {\bibfnamefont {H.~R.}\ \bibnamefont
  {Jiang}}, \bibinfo {author} {\bibfnamefont {N.}~\bibnamefont {Yoshinaga}}, \
  and\ \bibinfo {author} {\bibfnamefont {M.}~\bibnamefont {Sano}},\ }\href
  {\doibase 10.1103/PhysRevLett.105.268302} {\bibfield  {journal} {\bibinfo
  {journal} {Phys. Rev. Lett.}\ }\textbf {\bibinfo {volume} {105}},\ \bibinfo
  {pages} {268302} (\bibinfo {year} {2010})}\BibitemShut {NoStop}%
\bibitem [{\citenamefont {Palacci}\ \emph {et~al.}(2013)\citenamefont
  {Palacci}, \citenamefont {Sacanna}, \citenamefont {Steinberg}, \citenamefont
  {Pine},\ and\ \citenamefont {Chaikin}}]{Palacci2013}%
  \BibitemOpen
  \bibfield  {author} {\bibinfo {author} {\bibfnamefont {J.}~\bibnamefont
  {Palacci}}, \bibinfo {author} {\bibfnamefont {S.}~\bibnamefont {Sacanna}},
  \bibinfo {author} {\bibfnamefont {A.~P.}\ \bibnamefont {Steinberg}}, \bibinfo
  {author} {\bibfnamefont {D.~J.}\ \bibnamefont {Pine}}, \ and\ \bibinfo
  {author} {\bibfnamefont {P.~M.}\ \bibnamefont {Chaikin}},\ }\href {\doibase
  10.1126/science.1230020} {\bibfield  {journal} {\bibinfo  {journal}
  {Science}\ }\textbf {\bibinfo {volume} {339}},\ \bibinfo {pages} {936}
  (\bibinfo {year} {2013})}\BibitemShut {NoStop}%
\bibitem [{\citenamefont {Khadka}\ \emph {et~al.}(2018)\citenamefont {Khadka},
  \citenamefont {Holubec}, \citenamefont {Yang},\ and\ \citenamefont
  {Cichos}}]{Khadka2018}%
  \BibitemOpen
  \bibfield  {author} {\bibinfo {author} {\bibfnamefont {U.}~\bibnamefont
  {Khadka}}, \bibinfo {author} {\bibfnamefont {V.}~\bibnamefont {Holubec}},
  \bibinfo {author} {\bibfnamefont {H.}~\bibnamefont {Yang}}, \ and\ \bibinfo
  {author} {\bibfnamefont {F.}~\bibnamefont {Cichos}},\ }\href {\doibase
  10.1038/s41467-018-06445-1} {\bibfield  {journal} {\bibinfo  {journal} {Nat.
  Commun.}\ }\textbf {\bibinfo {volume} {9}},\ \bibinfo {pages} {3864}
  (\bibinfo {year} {2018})}\BibitemShut {NoStop}%
\bibitem [{\citenamefont {Ohgushi}\ \emph {et~al.}(2000)\citenamefont
  {Ohgushi}, \citenamefont {Murakami},\ and\ \citenamefont
  {Nagaosa}}]{Ohgushi2000}%
  \BibitemOpen
  \bibfield  {author} {\bibinfo {author} {\bibfnamefont {K.}~\bibnamefont
  {Ohgushi}}, \bibinfo {author} {\bibfnamefont {S.}~\bibnamefont {Murakami}}, \
  and\ \bibinfo {author} {\bibfnamefont {N.}~\bibnamefont {Nagaosa}},\ }\href
  {\doibase 10.1103/PhysRevB.62.R6065} {\bibfield  {journal} {\bibinfo
  {journal} {Phys. Rev. B}\ }\textbf {\bibinfo {volume} {62}},\ \bibinfo
  {pages} {R6065} (\bibinfo {year} {2000})}\BibitemShut {NoStop}%
\bibitem [{\citenamefont {Souslov}\ \emph {et~al.}(2017)\citenamefont
  {Souslov}, \citenamefont {{Van Zuiden}}, \citenamefont {Bartolo},\ and\
  \citenamefont {Vitelli}}]{Souslov2017}%
  \BibitemOpen
  \bibfield  {author} {\bibinfo {author} {\bibfnamefont {A.}~\bibnamefont
  {Souslov}}, \bibinfo {author} {\bibfnamefont {B.~C.}\ \bibnamefont {{Van
  Zuiden}}}, \bibinfo {author} {\bibfnamefont {D.}~\bibnamefont {Bartolo}}, \
  and\ \bibinfo {author} {\bibfnamefont {V.}~\bibnamefont {Vitelli}},\ }\href
  {\doibase 10.1038/nphys4193} {\bibfield  {journal} {\bibinfo  {journal} {Nat.
  Phys.}\ }\textbf {\bibinfo {volume} {13}},\ \bibinfo {pages} {1091} (\bibinfo
  {year} {2017})}\BibitemShut {NoStop}%
\bibitem [{\citenamefont {Shankar}\ \emph {et~al.}(2017)\citenamefont
  {Shankar}, \citenamefont {Bowick},\ and\ \citenamefont
  {Marchetti}}]{Shankar2017}%
  \BibitemOpen
  \bibfield  {author} {\bibinfo {author} {\bibfnamefont {S.}~\bibnamefont
  {Shankar}}, \bibinfo {author} {\bibfnamefont {M.~J.}\ \bibnamefont {Bowick}},
  \ and\ \bibinfo {author} {\bibfnamefont {M.~C.}\ \bibnamefont {Marchetti}},\
  }\href {\doibase 10.1103/PhysRevX.7.031039} {\bibfield  {journal} {\bibinfo
  {journal} {Phys. Rev. X}\ }\textbf {\bibinfo {volume} {7}},\ \bibinfo {pages}
  {031039} (\bibinfo {year} {2017})}\BibitemShut {NoStop}%
\bibitem [{\citenamefont {Souslov}\ \emph {et~al.}(2019)\citenamefont
  {Souslov}, \citenamefont {Dasbiswas}, \citenamefont {Fruchart}, \citenamefont
  {Vaikuntanathan},\ and\ \citenamefont {Vitelli}}]{Souslov2018}%
  \BibitemOpen
  \bibfield  {author} {\bibinfo {author} {\bibfnamefont {A.}~\bibnamefont
  {Souslov}}, \bibinfo {author} {\bibfnamefont {K.}~\bibnamefont {Dasbiswas}},
  \bibinfo {author} {\bibfnamefont {M.}~\bibnamefont {Fruchart}}, \bibinfo
  {author} {\bibfnamefont {S.}~\bibnamefont {Vaikuntanathan}}, \ and\ \bibinfo
  {author} {\bibfnamefont {V.}~\bibnamefont {Vitelli}},\ }\href {\doibase
  10.1103/PhysRevLett.122.128001} {\bibfield  {journal} {\bibinfo  {journal}
  {Phys. Rev. Lett.}\ }\textbf {\bibinfo {volume} {122}},\ \bibinfo {pages}
  {128001} (\bibinfo {year} {2019})}\BibitemShut {NoStop}%
\bibitem [{\citenamefont {Klitzing}\ \emph {et~al.}(1980)\citenamefont
  {Klitzing}, \citenamefont {Dorda},\ and\ \citenamefont
  {Pepper}}]{Klitzing1980}%
  \BibitemOpen
  \bibfield  {author} {\bibinfo {author} {\bibfnamefont {K.~v.}\ \bibnamefont
  {Klitzing}}, \bibinfo {author} {\bibfnamefont {G.}~\bibnamefont {Dorda}}, \
  and\ \bibinfo {author} {\bibfnamefont {M.}~\bibnamefont {Pepper}},\ }\href
  {\doibase 10.1103/PhysRevLett.45.494} {\bibfield  {journal} {\bibinfo
  {journal} {Phys. Rev. Lett.}\ }\textbf {\bibinfo {volume} {45}},\ \bibinfo
  {pages} {494} (\bibinfo {year} {1980})}\BibitemShut {NoStop}%
\bibitem [{\citenamefont {Sokolov}\ \emph {et~al.}(2007)\citenamefont
  {Sokolov}, \citenamefont {Aranson}, \citenamefont {Kessler},\ and\
  \citenamefont {Goldstein}}]{Sokolov2007}%
  \BibitemOpen
  \bibfield  {author} {\bibinfo {author} {\bibfnamefont {A.}~\bibnamefont
  {Sokolov}}, \bibinfo {author} {\bibfnamefont {I.~S.}\ \bibnamefont
  {Aranson}}, \bibinfo {author} {\bibfnamefont {J.~O.}\ \bibnamefont
  {Kessler}}, \ and\ \bibinfo {author} {\bibfnamefont {R.~E.}\ \bibnamefont
  {Goldstein}},\ }\href {\doibase 10.1103/PhysRevLett.98.158102} {\bibfield
  {journal} {\bibinfo  {journal} {Phys. Rev. Lett.}\ }\textbf {\bibinfo
  {volume} {98}},\ \bibinfo {pages} {158102} (\bibinfo {year}
  {2007})}\BibitemShut {NoStop}%
\bibitem [{\citenamefont {Cavagna}\ \emph {et~al.}(2010)\citenamefont
  {Cavagna}, \citenamefont {Cimarelli}, \citenamefont {Giardina}, \citenamefont
  {Parisi}, \citenamefont {Santagati}, \citenamefont {Stefanini},\ and\
  \citenamefont {Viale}}]{Cavagna2010}%
  \BibitemOpen
  \bibfield  {author} {\bibinfo {author} {\bibfnamefont {A.}~\bibnamefont
  {Cavagna}}, \bibinfo {author} {\bibfnamefont {A.}~\bibnamefont {Cimarelli}},
  \bibinfo {author} {\bibfnamefont {I.}~\bibnamefont {Giardina}}, \bibinfo
  {author} {\bibfnamefont {G.}~\bibnamefont {Parisi}}, \bibinfo {author}
  {\bibfnamefont {R.}~\bibnamefont {Santagati}}, \bibinfo {author}
  {\bibfnamefont {F.}~\bibnamefont {Stefanini}}, \ and\ \bibinfo {author}
  {\bibfnamefont {M.}~\bibnamefont {Viale}},\ }\href {\doibase
  10.1073/pnas.1005766107} {\bibfield  {journal} {\bibinfo  {journal} {Proc.
  Natl. Acad. Sci. U.S.A.}\ }\textbf {\bibinfo {volume} {107}},\ \bibinfo
  {pages} {11865} (\bibinfo {year} {2010})}\BibitemShut {NoStop}%
\bibitem [{\citenamefont {Deseigne}\ \emph {et~al.}(2010)\citenamefont
  {Deseigne}, \citenamefont {Dauchot},\ and\ \citenamefont
  {Chat{\'{e}}}}]{Deseigne2010}%
  \BibitemOpen
  \bibfield  {author} {\bibinfo {author} {\bibfnamefont {J.}~\bibnamefont
  {Deseigne}}, \bibinfo {author} {\bibfnamefont {O.}~\bibnamefont {Dauchot}}, \
  and\ \bibinfo {author} {\bibfnamefont {H.}~\bibnamefont {Chat{\'{e}}}},\
  }\href {\doibase 10.1103/PhysRevLett.105.098001} {\bibfield  {journal}
  {\bibinfo  {journal} {Phys. Rev. Lett.}\ }\textbf {\bibinfo {volume} {105}},\
  \bibinfo {pages} {098001} (\bibinfo {year} {2010})}\BibitemShut {NoStop}%
\bibitem [{\citenamefont {Schaller}\ \emph {et~al.}(2010)\citenamefont
  {Schaller}, \citenamefont {Weber}, \citenamefont {Semmrich}, \citenamefont
  {Frey},\ and\ \citenamefont {Bausch}}]{Schaller2010}%
  \BibitemOpen
  \bibfield  {author} {\bibinfo {author} {\bibfnamefont {V.}~\bibnamefont
  {Schaller}}, \bibinfo {author} {\bibfnamefont {C.}~\bibnamefont {Weber}},
  \bibinfo {author} {\bibfnamefont {C.}~\bibnamefont {Semmrich}}, \bibinfo
  {author} {\bibfnamefont {E.}~\bibnamefont {Frey}}, \ and\ \bibinfo {author}
  {\bibfnamefont {A.~R.}\ \bibnamefont {Bausch}},\ }\href {\doibase
  10.1038/nature09312} {\bibfield  {journal} {\bibinfo  {journal} {Nature}\
  }\textbf {\bibinfo {volume} {467}},\ \bibinfo {pages} {73} (\bibinfo {year}
  {2010})}\BibitemShut {NoStop}%
\bibitem [{\citenamefont {Nishiguchi}\ \emph {et~al.}(2018)\citenamefont
  {Nishiguchi}, \citenamefont {Aranson}, \citenamefont {Snezhko},\ and\
  \citenamefont {Sokolov}}]{Nishiguchi2018}%
  \BibitemOpen
  \bibfield  {author} {\bibinfo {author} {\bibfnamefont {D.}~\bibnamefont
  {Nishiguchi}}, \bibinfo {author} {\bibfnamefont {I.~S.}\ \bibnamefont
  {Aranson}}, \bibinfo {author} {\bibfnamefont {A.}~\bibnamefont {Snezhko}}, \
  and\ \bibinfo {author} {\bibfnamefont {A.}~\bibnamefont {Sokolov}},\ }\href
  {\doibase 10.1038/s41467-018-06842-6} {\bibfield  {journal} {\bibinfo
  {journal} {Nat. Commun.}\ }\textbf {\bibinfo {volume} {9}},\ \bibinfo {pages}
  {4486} (\bibinfo {year} {2018})}\BibitemShut {NoStop}%
\bibitem [{\citenamefont {Toner}\ and\ \citenamefont {Tu}(1995)}]{Toner1995}%
  \BibitemOpen
  \bibfield  {author} {\bibinfo {author} {\bibfnamefont {J.}~\bibnamefont
  {Toner}}\ and\ \bibinfo {author} {\bibfnamefont {Y.}~\bibnamefont {Tu}},\
  }\href {\doibase 10.1103/PhysRevLett.75.4326} {\bibfield  {journal} {\bibinfo
   {journal} {Phys. Rev. Lett.}\ }\textbf {\bibinfo {volume} {75}},\ \bibinfo
  {pages} {4326} (\bibinfo {year} {1995})}\BibitemShut {NoStop}%
\bibitem [{\citenamefont {Toner}\ and\ \citenamefont {Tu}(1998)}]{Toner1998}%
  \BibitemOpen
  \bibfield  {author} {\bibinfo {author} {\bibfnamefont {J.}~\bibnamefont
  {Toner}}\ and\ \bibinfo {author} {\bibfnamefont {Y.}~\bibnamefont {Tu}},\
  }\href {\doibase 10.1103/PhysRevE.58.4828} {\bibfield  {journal} {\bibinfo
  {journal} {Phys. Rev. E}\ }\textbf {\bibinfo {volume} {58}},\ \bibinfo
  {pages} {4828} (\bibinfo {year} {1998})}\BibitemShut {NoStop}%
\bibitem [{\citenamefont {Toner}\ \emph {et~al.}(2005)\citenamefont {Toner},
  \citenamefont {Tu},\ and\ \citenamefont {Ramaswamy}}]{Toner2005}%
  \BibitemOpen
  \bibfield  {author} {\bibinfo {author} {\bibfnamefont {J.}~\bibnamefont
  {Toner}}, \bibinfo {author} {\bibfnamefont {Y.}~\bibnamefont {Tu}}, \ and\
  \bibinfo {author} {\bibfnamefont {S.}~\bibnamefont {Ramaswamy}},\ }\href
  {\doibase 10.1016/j.aop.2005.04.011} {\bibfield  {journal} {\bibinfo
  {journal} {Ann. Phys.}\ }\textbf {\bibinfo {volume} {318}},\ \bibinfo {pages}
  {170} (\bibinfo {year} {2005})}\BibitemShut {NoStop}%
\bibitem [{\citenamefont {Marchetti}\ \emph {et~al.}(2013)\citenamefont
  {Marchetti}, \citenamefont {Joanny}, \citenamefont {Ramaswamy}, \citenamefont
  {Liverpool}, \citenamefont {Prost}, \citenamefont {Rao},\ and\ \citenamefont
  {Simha}}]{Marchetti2013}%
  \BibitemOpen
  \bibfield  {author} {\bibinfo {author} {\bibfnamefont {M.~C.}\ \bibnamefont
  {Marchetti}}, \bibinfo {author} {\bibfnamefont {J.~F.}\ \bibnamefont
  {Joanny}}, \bibinfo {author} {\bibfnamefont {S.}~\bibnamefont {Ramaswamy}},
  \bibinfo {author} {\bibfnamefont {T.~B.}\ \bibnamefont {Liverpool}}, \bibinfo
  {author} {\bibfnamefont {J.}~\bibnamefont {Prost}}, \bibinfo {author}
  {\bibfnamefont {M.}~\bibnamefont {Rao}}, \ and\ \bibinfo {author}
  {\bibfnamefont {R.~A.}\ \bibnamefont {Simha}},\ }\href {\doibase
  10.1103/RevModPhys.85.1143} {\bibfield  {journal} {\bibinfo  {journal} {Rev.
  Mod. Phys.}\ }\textbf {\bibinfo {volume} {85}},\ \bibinfo {pages} {1143}
  (\bibinfo {year} {2013})}\BibitemShut {NoStop}%
\bibitem [{\citenamefont {Bertin}\ \emph {et~al.}(2006)\citenamefont {Bertin},
  \citenamefont {Droz},\ and\ \citenamefont {Gr{\'{e}}goire}}]{Bertin2006}%
  \BibitemOpen
  \bibfield  {author} {\bibinfo {author} {\bibfnamefont {E.}~\bibnamefont
  {Bertin}}, \bibinfo {author} {\bibfnamefont {M.}~\bibnamefont {Droz}}, \ and\
  \bibinfo {author} {\bibfnamefont {G.}~\bibnamefont {Gr{\'{e}}goire}},\ }\href
  {\doibase 10.1103/PhysRevE.74.022101} {\bibfield  {journal} {\bibinfo
  {journal} {Phys. Rev. E}\ }\textbf {\bibinfo {volume} {74}},\ \bibinfo
  {pages} {022101} (\bibinfo {year} {2006})}\BibitemShut {NoStop}%
\bibitem [{\citenamefont {Peshkov}\ \emph {et~al.}(2014)\citenamefont
  {Peshkov}, \citenamefont {Bertin}, \citenamefont {F.Ginelli},\ and\
  \citenamefont {Chat{\'e}}}]{Peshkov2014}%
  \BibitemOpen
  \bibfield  {author} {\bibinfo {author} {\bibfnamefont {A.}~\bibnamefont
  {Peshkov}}, \bibinfo {author} {\bibfnamefont {E.}~\bibnamefont {Bertin}},
  \bibinfo {author} {\bibnamefont {F.Ginelli}}, \ and\ \bibinfo {author}
  {\bibfnamefont {H.}~\bibnamefont {Chat{\'e}}},\ }\href {\doibase
  10.1140/epjst/e2014-02193-y} {\bibfield  {journal} {\bibinfo  {journal} {Eur.
  Phys. J. Spec. Top.}\ }\textbf {\bibinfo {volume} {223}},\ \bibinfo {pages}
  {1315} (\bibinfo {year} {2014})}\BibitemShut {NoStop}%
\bibitem [{\citenamefont {Farrell}\ \emph {et~al.}(2012)\citenamefont
  {Farrell}, \citenamefont {Marchetti}, \citenamefont {Marenduzzo},\ and\
  \citenamefont {Tailleur}}]{Farrell2012}%
  \BibitemOpen
  \bibfield  {author} {\bibinfo {author} {\bibfnamefont {F.~D.}\ \bibnamefont
  {Farrell}}, \bibinfo {author} {\bibfnamefont {M.~C.}\ \bibnamefont
  {Marchetti}}, \bibinfo {author} {\bibfnamefont {D.}~\bibnamefont
  {Marenduzzo}}, \ and\ \bibinfo {author} {\bibfnamefont {J.}~\bibnamefont
  {Tailleur}},\ }\href {\doibase 10.1103/PhysRevLett.108.248101} {\bibfield
  {journal} {\bibinfo  {journal} {Phys. Rev. Lett.}\ }\textbf {\bibinfo
  {volume} {108}},\ \bibinfo {pages} {248101} (\bibinfo {year}
  {2012})}\BibitemShut {NoStop}%
\bibitem [{\citenamefont {Solon}\ and\ \citenamefont
  {Tailleur}(2013)}]{Solon2013}%
  \BibitemOpen
  \bibfield  {author} {\bibinfo {author} {\bibfnamefont {A.~P.}\ \bibnamefont
  {Solon}}\ and\ \bibinfo {author} {\bibfnamefont {J.}~\bibnamefont
  {Tailleur}},\ }\href {\doibase 10.1103/PhysRevLett.111.078101} {\bibfield
  {journal} {\bibinfo  {journal} {Phys. Rev. Lett.}\ }\textbf {\bibinfo
  {volume} {111}},\ \bibinfo {pages} {078101} (\bibinfo {year}
  {2013})}\BibitemShut {NoStop}%
\bibitem [{\citenamefont {Suzuki}\ \emph {et~al.}(2015)\citenamefont {Suzuki},
  \citenamefont {Weber}, \citenamefont {Frey},\ and\ \citenamefont
  {Bausch}}]{Suzuki2015}%
  \BibitemOpen
  \bibfield  {author} {\bibinfo {author} {\bibfnamefont {R.}~\bibnamefont
  {Suzuki}}, \bibinfo {author} {\bibfnamefont {C.~A.}\ \bibnamefont {Weber}},
  \bibinfo {author} {\bibfnamefont {E.}~\bibnamefont {Frey}}, \ and\ \bibinfo
  {author} {\bibfnamefont {A.~R.}\ \bibnamefont {Bausch}},\ }\href {\doibase
  10.1038/nphys3423} {\bibfield  {journal} {\bibinfo  {journal} {Nat. Phys.}\
  }\textbf {\bibinfo {volume} {11}},\ \bibinfo {pages} {839} (\bibinfo {year}
  {2015})}\BibitemShut {NoStop}%
\bibitem [{\citenamefont {Bricard}\ \emph {et~al.}(2013)\citenamefont
  {Bricard}, \citenamefont {Caussin}, \citenamefont {Desreumaux}, \citenamefont
  {Dauchot},\ and\ \citenamefont {Bartolo}}]{Bricard2013}%
  \BibitemOpen
  \bibfield  {author} {\bibinfo {author} {\bibfnamefont {A.}~\bibnamefont
  {Bricard}}, \bibinfo {author} {\bibfnamefont {J.~B.}\ \bibnamefont
  {Caussin}}, \bibinfo {author} {\bibfnamefont {N.}~\bibnamefont {Desreumaux}},
  \bibinfo {author} {\bibfnamefont {O.}~\bibnamefont {Dauchot}}, \ and\
  \bibinfo {author} {\bibfnamefont {D.}~\bibnamefont {Bartolo}},\ }\href
  {\doibase 10.1038/nature12673} {\bibfield  {journal} {\bibinfo  {journal}
  {Nature}\ }\textbf {\bibinfo {volume} {503}},\ \bibinfo {pages} {95}
  (\bibinfo {year} {2013})}\BibitemShut {NoStop}%
\bibitem [{\citenamefont {Saragosti}\ \emph {et~al.}(2011)\citenamefont
  {Saragosti}, \citenamefont {Calvez}, \citenamefont {Bournaveas},
  \citenamefont {Perthame}, \citenamefont {Buguin},\ and\ \citenamefont
  {Silberzan}}]{Saragosti2011}%
  \BibitemOpen
  \bibfield  {author} {\bibinfo {author} {\bibfnamefont {J.}~\bibnamefont
  {Saragosti}}, \bibinfo {author} {\bibfnamefont {V.}~\bibnamefont {Calvez}},
  \bibinfo {author} {\bibfnamefont {N.}~\bibnamefont {Bournaveas}}, \bibinfo
  {author} {\bibfnamefont {B.}~\bibnamefont {Perthame}}, \bibinfo {author}
  {\bibfnamefont {A.}~\bibnamefont {Buguin}}, \ and\ \bibinfo {author}
  {\bibfnamefont {P.}~\bibnamefont {Silberzan}},\ }\href {\doibase
  10.1073/pnas.1101996108} {\bibfield  {journal} {\bibinfo  {journal} {Proc.
  Natl. Acad. Sci. U.S.A.}\ }\textbf {\bibinfo {volume} {108}},\ \bibinfo
  {pages} {16235} (\bibinfo {year} {2011})}\BibitemShut {NoStop}%
\bibitem [{Note1()}]{Note1}%
  \BibitemOpen
  \bibinfo {note} {We assume that, while these pillars can influence the flow
  of particles, their sizes are small enough such that the conditions discussed
  above Eq.~\protect \textup {\hbox {\mathsurround \z@ \protect \normalfont
  (\ignorespaces \ref {eq:stokes}\unskip \@@italiccorr )}} are
  satisfied.}\BibitemShut {Stop}%
\bibitem [{Note2()}]{Note2}%
  \BibitemOpen
  \bibinfo {note} {The reversed flow $-\protect \mathbf {v}_{\protect \rm ss}$
  is also allowed as a steady-state solution; the initial condition determines
  which flow can be realized in a system.}\BibitemShut {Stop}%
\bibitem [{\citenamefont {Wioland}\ \emph {et~al.}(2016)\citenamefont
  {Wioland}, \citenamefont {Woodhouse}, \citenamefont {Dunkel},\ and\
  \citenamefont {Goldstein}}]{Wioland2016}%
  \BibitemOpen
  \bibfield  {author} {\bibinfo {author} {\bibfnamefont {H.}~\bibnamefont
  {Wioland}}, \bibinfo {author} {\bibfnamefont {F.~G.}\ \bibnamefont
  {Woodhouse}}, \bibinfo {author} {\bibfnamefont {J.}~\bibnamefont {Dunkel}}, \
  and\ \bibinfo {author} {\bibfnamefont {R.~E.}\ \bibnamefont {Goldstein}},\
  }\href {\doibase 10.1038/nphys3607} {\bibfield  {journal} {\bibinfo
  {journal} {Nat. Phys.}\ }\textbf {\bibinfo {volume} {12}},\ \bibinfo {pages}
  {341} (\bibinfo {year} {2016})}\BibitemShut {NoStop}%
\bibitem [{\citenamefont {Pearce}\ and\ \citenamefont
  {Turner}(2015)}]{Pearce2015}%
  \BibitemOpen
  \bibfield  {author} {\bibinfo {author} {\bibfnamefont {D.~J.}\ \bibnamefont
  {Pearce}}\ and\ \bibinfo {author} {\bibfnamefont {M.~S.}\ \bibnamefont
  {Turner}},\ }\href {\doibase 10.1098/rsif.2015.0520} {\bibfield  {journal}
  {\bibinfo  {journal} {J. Royal Soc. Interface}\ }\textbf {\bibinfo {volume}
  {12}},\ \bibinfo {pages} {20150520} (\bibinfo {year} {2015})}\BibitemShut
  {NoStop}%
\bibitem [{\citenamefont {Smith}(1985)}]{Smith1985}%
  \BibitemOpen
  \bibfield  {author} {\bibinfo {author} {\bibfnamefont {G.~D.}\ \bibnamefont
  {Smith}},\ }\href@noop {} {\emph {\bibinfo {title} {Numerical solution of
  partial differential equations : finite difference methods}}}\ (\bibinfo
  {publisher} {Oxford University Press},\ \bibinfo {year} {1985})\BibitemShut
  {NoStop}%
\bibitem [{\citenamefont {Fukui}\ \emph {et~al.}(2005)\citenamefont {Fukui},
  \citenamefont {Hatsugai},\ and\ \citenamefont {Suzuki}}]{Fukui2005}%
  \BibitemOpen
  \bibfield  {author} {\bibinfo {author} {\bibfnamefont {T.}~\bibnamefont
  {Fukui}}, \bibinfo {author} {\bibfnamefont {Y.}~\bibnamefont {Hatsugai}}, \
  and\ \bibinfo {author} {\bibfnamefont {H.}~\bibnamefont {Suzuki}},\ }\href
  {\doibase 10.1143/JPSJ.74.1674} {\bibfield  {journal} {\bibinfo  {journal}
  {J. Phys. Soc. of Jpn.}\ }\textbf {\bibinfo {volume} {74}},\ \bibinfo {pages}
  {1674} (\bibinfo {year} {2005})}\BibitemShut {NoStop}%
\bibitem [{\citenamefont {Silveirinha}(2019)}]{Silveirinha2019}%
  \BibitemOpen
  \bibfield  {author} {\bibinfo {author} {\bibfnamefont {M.~G.}\ \bibnamefont
  {Silveirinha}},\ }\href {\doibase 10.1103/PhysRevX.9.011037} {\bibfield
  {journal} {\bibinfo  {journal} {Phys. Rev. X}\ }\textbf {\bibinfo {volume}
  {9}},\ \bibinfo {pages} {011037} (\bibinfo {year} {2019})}\BibitemShut
  {NoStop}%
\bibitem [{\citenamefont {Lee}(2016)}]{Lee2016}%
  \BibitemOpen
  \bibfield  {author} {\bibinfo {author} {\bibfnamefont {T.~E.}\ \bibnamefont
  {Lee}},\ }\href {\doibase 10.1103/PhysRevLett.116.133903} {\bibfield
  {journal} {\bibinfo  {journal} {Phys. Rev. Lett.}\ }\textbf {\bibinfo
  {volume} {116}},\ \bibinfo {pages} {133903} (\bibinfo {year}
  {2016})}\BibitemShut {NoStop}%
\bibitem [{\citenamefont {Leykam}\ \emph {et~al.}(2017)\citenamefont {Leykam},
  \citenamefont {Bliokh}, \citenamefont {Huang}, \citenamefont {Chong},\ and\
  \citenamefont {Nori}}]{Leykam2017}%
  \BibitemOpen
  \bibfield  {author} {\bibinfo {author} {\bibfnamefont {D.}~\bibnamefont
  {Leykam}}, \bibinfo {author} {\bibfnamefont {K.~Y.}\ \bibnamefont {Bliokh}},
  \bibinfo {author} {\bibfnamefont {C.}~\bibnamefont {Huang}}, \bibinfo
  {author} {\bibfnamefont {Y.~D.}\ \bibnamefont {Chong}}, \ and\ \bibinfo
  {author} {\bibfnamefont {F.}~\bibnamefont {Nori}},\ }\href {\doibase
  10.1103/PhysRevLett.118.040401} {\bibfield  {journal} {\bibinfo  {journal}
  {Phys. Rev. Lett.}\ }\textbf {\bibinfo {volume} {118}},\ \bibinfo {pages}
  {040401} (\bibinfo {year} {2017})}\BibitemShut {NoStop}%
\bibitem [{\citenamefont {Kunst}\ \emph {et~al.}(2018)\citenamefont {Kunst},
  \citenamefont {Edvardsson}, \citenamefont {Budich},\ and\ \citenamefont
  {Bergholtz}}]{Kunst2018}%
  \BibitemOpen
  \bibfield  {author} {\bibinfo {author} {\bibfnamefont {F.~K.}\ \bibnamefont
  {Kunst}}, \bibinfo {author} {\bibfnamefont {E.}~\bibnamefont {Edvardsson}},
  \bibinfo {author} {\bibfnamefont {J.~C.}\ \bibnamefont {Budich}}, \ and\
  \bibinfo {author} {\bibfnamefont {E.~J.}\ \bibnamefont {Bergholtz}},\ }\href
  {\doibase 10.1103/PhysRevLett.121.026808} {\bibfield  {journal} {\bibinfo
  {journal} {Phys. Rev. Lett.}\ }\textbf {\bibinfo {volume} {121}},\ \bibinfo
  {pages} {026808} (\bibinfo {year} {2018})}\BibitemShut {NoStop}%
\bibitem [{\citenamefont {Yao}\ and\ \citenamefont {Wang}(2018)}]{Yao2018}%
  \BibitemOpen
  \bibfield  {author} {\bibinfo {author} {\bibfnamefont {S.}~\bibnamefont
  {Yao}}\ and\ \bibinfo {author} {\bibfnamefont {Z.}~\bibnamefont {Wang}},\
  }\href {\doibase 10.1103/PhysRevLett.121.086803} {\bibfield  {journal}
  {\bibinfo  {journal} {Phys. Rev. Lett.}\ }\textbf {\bibinfo {volume} {121}},\
  \bibinfo {pages} {086803} (\bibinfo {year} {2018})}\BibitemShut {NoStop}%
\bibitem [{\citenamefont {Gong}\ \emph {et~al.}(2018)\citenamefont {Gong},
  \citenamefont {Ashida}, \citenamefont {Kawabata}, \citenamefont {Takasan},
  \citenamefont {Higashikawa},\ and\ \citenamefont {Ueda}}]{Gong2018}%
  \BibitemOpen
  \bibfield  {author} {\bibinfo {author} {\bibfnamefont {Z.}~\bibnamefont
  {Gong}}, \bibinfo {author} {\bibfnamefont {Y.}~\bibnamefont {Ashida}},
  \bibinfo {author} {\bibfnamefont {K.}~\bibnamefont {Kawabata}}, \bibinfo
  {author} {\bibfnamefont {K.}~\bibnamefont {Takasan}}, \bibinfo {author}
  {\bibfnamefont {S.}~\bibnamefont {Higashikawa}}, \ and\ \bibinfo {author}
  {\bibfnamefont {M.}~\bibnamefont {Ueda}},\ }\href {\doibase
  10.1103/PhysRevX.8.031079} {\bibfield  {journal} {\bibinfo  {journal} {Phys.
  Rev. X}\ }\textbf {\bibinfo {volume} {8}},\ \bibinfo {pages} {031079}
  (\bibinfo {year} {2018})}\BibitemShut {NoStop}%
\bibitem [{\citenamefont {Longhi}(2019)}]{Longhi2019}%
  \BibitemOpen
  \bibfield  {author} {\bibinfo {author} {\bibfnamefont {S.}~\bibnamefont
  {Longhi}},\ }\href {\doibase 10.1103/PhysRevResearch.1.023013} {\bibfield
  {journal} {\bibinfo  {journal} {Phys. Rev. Res.}\ }\textbf {\bibinfo {volume}
  {1}},\ \bibinfo {pages} {023013} (\bibinfo {year} {2019})}\BibitemShut
  {NoStop}%
\end{thebibliography}%

\widetext
\pagebreak
\begin{center}
\textbf{\large Supplementary Materials}
\end{center}

\renewcommand{\theequation}{S\arabic{equation}}
\renewcommand{\thefigure}{S\arabic{figure}}
\renewcommand{\bibnumfmt}[1]{[S#1]}
\setcounter{equation}{0}
\setcounter{figure}{0}

\subsection{The spatial variation of the density field and the steady-state flow}
The equations for a steady state read as
\begin{equation}
 \rho_{\rm ss} \nabla \cdot \mathbf{v}_{\rm ss} + (\nabla \rho_{\rm ss}) \cdot \mathbf{v}_{\rm ss} = 0,
\label{simple-toner-tu-eq-steady1}
\end{equation}
\begin{equation}
\lambda (\mathbf{v}_{\rm ss} \cdot \nabla) \mathbf{v}_{\rm ss} = -\frac{c^2}{\rho_{\rm ss}} \nabla \rho_{\rm ss}.
\label{simple-toner-tu-eq-steady2}
\end{equation}
Equation~\eqref{simple-toner-tu-eq-steady2} implies that $a \nabla \rho_{\rm ss} / \rho_{\rm ss} = \mathcal{O}((|\mathbf{v}_{\rm ss}|/c)^2)$, where $a$ is the length of each side of the unit cell. Therefore, when $|\mathbf{v}_{\rm ss}|/c =|\mathbf{v}'_{\rm ss}|$ is small, one can ignore the spatial variation of the density field in a steady state. We thus obtain  $a\nabla \cdot \mathbf{v}_{\rm ss}/c = \mathcal{O}((|\mathbf{v}_{\rm ss}|/c)^3)$ from Eq.~\eqref{simple-toner-tu-eq-steady1}. This implies that the divergence of the steady-state flow can also be ignored and the Toner-Tu equations permit the steady-state flow whose divergence is zero. 

\begin{figure}[b]
  \includegraphics[width=84mm, bb=0 0 650 420,clip]{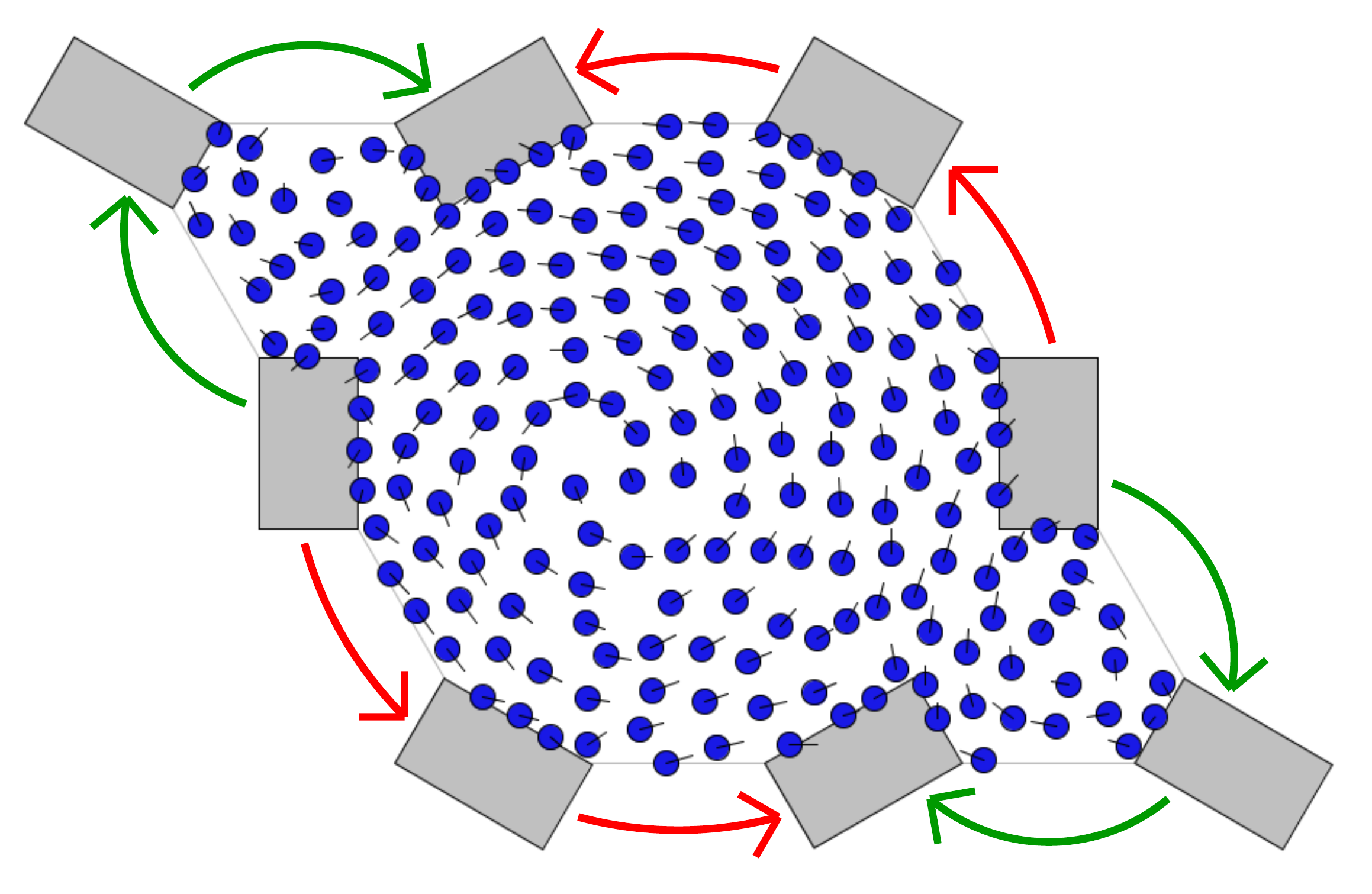}
\caption{\label{fig:particle-simulation} Typical snapshot from the particle-based simulation for our periodic kagome lattice. Each circle represents an active particle and a bar extending from it indicates the velocity of each particle. The total number of particles is 200 and the parameters are $m=\gamma=1$, $F_0=0.5$, $a=60$, $b=5$, $\kappa=0.1$, $k_p=100$, and $k_{\rm B}T=2\times 10^{-4}$.}
\end{figure}

To confirm the validity of the steady-state flow considered (cf. Fig.~1(b) in the main text), we perform the particle-based simulation for the bulk system. We use the following particle model~\cite{Souslov2017}, 
\eqn{
m\dot{\mathbf{v}}_i &=& -\gamma \mathbf{v}_i + F_0 \left( \hat{\mathbf{v}}_i + \sum_{\langle \mathbf{x}_i,\mathbf{x}_j \rangle} \hat{\mathbf{v}}_i/N \right)\nonumber\\
&{}& + \sum_k \nabla_i U(|\mathbf{x}_i-\mathbf{x}_k|) + \sqrt{2\gamma k_B T} \hat{\zeta}_i (t),
\label{particle-model}
}
where $\mathbf{x}_i$ and $\mathbf{v}_i$ are the position and velocity of each particle and $\hat{\mathbf{v}}_i$ is the unit vector parallel to $\mathbf{v}_i$, $\langle \mathbf{x}_i,\mathbf{x}_j \rangle$ denotes the neighbors of the $i$-th particle and $N$ is the number of neighboring particles, $U(|\mathbf{x}_i-\mathbf{x}_k|)$ represents repulsive potential between particles introduced to avoid the condensation, and $\sqrt{2\gamma k_B T} \hat{\zeta}_i (t)$ is the white Gaussian noise term. Here, we use Yukawa potential $U(r) = b/(re^{\kappa r})$. We implement the pillars by using a steep one-sided harmonic repulsive potential $k_p x^2/2$. Performing the numerical simulation of this model in our periodic kagome lattice starting from the random initial state, we observe the emergence of the steady-state flow discussed in the main text (see the Supplementary Movie published with the manuscript). Supplementary Figure~\ref{fig:particle-simulation} shows a typical snapshot of the particle-based simulation. 
These results justify the presence of the steady-state flow in Fig.~1(b) in the main text as the present particle-based model should be described by the Toner-Tu equations in the hydrodynamic limit.

We can also make a qualitative discussion about the emergence of the steady-state flow. Since the actual system must be finite and have walls around the system, the flow cannot come in and out of the edge of the entire system. Assuming the periodicity of the steady-state flow imposed by the periodic structure of the bulk, one can conclude that flows in and out of each unit cell are also forbidden in the bulk of the system (cf. the red arrows in the center of Supplementary Figure~\ref{fig:qualitative-discussion}). Then, around the boundary of a unit cell, the steady-state flow must be parallel to the edge of a unit cell. For these reasons, the flow in Fig.~1(b) in the main text can naturally emerge as a steady state. 

We note that the linear stability analysis is a standard method to confirm the validity of the steady state. Our analysis based on the linearized Toner-Tu equations corresponds to the linear stability analysis around the steady state of interest here. In our setup,  since the effective Hamiltonian is found to be Hermitian and thus exhibits real eigenvalues, unstable modes are absent at the level of the current analysis. We note that,  while the above arguments can justify the presence and stability of the steady-state flow considered in the main text, the uniqueness of the steady state still remains as an open question and should be addressed by performing full nonlinear calculations of the Toner-Tu equations. 

\begin{figure}[t]
  \includegraphics[width=84mm, bb=0 0 600 310,clip]{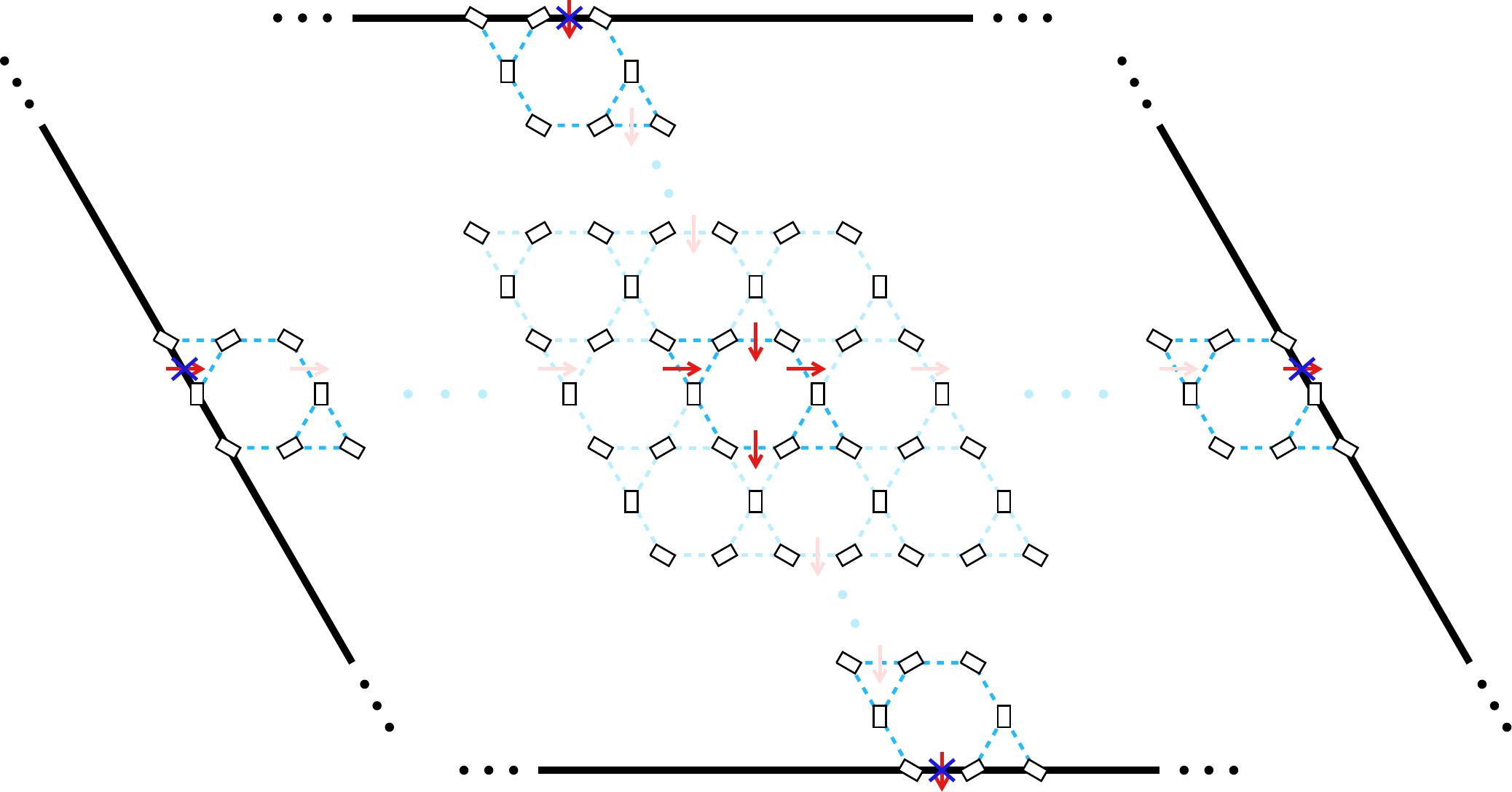}
\caption{\label{fig:qualitative-discussion} Schematic diagram of the qualitative discussions for the steady-state flow. The flow depicted by the red arrows is prohibited by the boundary conditions.}
\end{figure}

\subsection{Linearization of the Toner-Tu equations and the derivation of the Schr\"{o}dinger-like equation}
Here we show the details about the derivation of the linearized equation. The full Toner-Tu equations read
\eqn{
&&\partial_t \rho + \nabla \cdot (\rho \mathbf{v}) = 0,\\
\label{toner-tu-eq-app1}
 \partial_t \mathbf{v} + \lambda (\mathbf{v} \cdot \nabla) \mathbf{v} &+& \lambda_2 (\nabla \cdot \mathbf{v}) \mathbf{v} + \lambda_3  \nabla |\mathbf{v}|^2
  \nonumber\\
= (\alpha-\beta|\mathbf{v}|^2)\mathbf{v}-\nabla P &+&D_B \nabla (\nabla \cdot \mathbf{v}) + D_T \nabla^2 \mathbf{v}+ D_2(\mathbf{v} \cdot \nabla)^2 \mathbf{v} + \mathbf{f}.
\label{toner-tu-eq-app2}
}
Since we neglect the diffusive terms and $\lambda_2,\lambda_3$ terms, we obtain the simplified Toner-Tu equations 
\eqn{
\partial_t \rho &+& \nabla \cdot (\rho \mathbf{v}) = 0,\\
\label{simple-toner-tu-eq1}
\partial_t \mathbf{v} + \lambda (\mathbf{v} \cdot \nabla) \mathbf{v} &=& (\alpha-\beta|\mathbf{v}|^2)\mathbf{v}-\frac{c^2}{\rho_{\rm ss}} \nabla \rho,
\label{simple-toner-tu-eq2}
}
where we assume the equation of state $P=c^2\rho/\rho_{\rm ss}$ with $c$ being the sound speed and that the external force $\mathbf{f}$ is absent. We consider the fluctuations of  density and velocity fields from a steady state, $\delta\rho({\bf r},t) = \rho({\bf r},t) - \rho_{\rm ss}({\bf r})$, $\delta\mathbf{v}({\bf r},t)=\mathbf{v}({\bf r},t)-\mathbf{v}_{\rm ss}({\bf r})$ and neglect their second-order contributions in Eq.~\eqref{simple-toner-tu-eq2}. Then, we obtain 
\begin{eqnarray}
\partial_t \delta \rho + (\mathbf{v}_{\rm ss} \cdot \nabla)\delta \rho &=& -\rho_{\rm ss} \nabla \cdot \delta \mathbf{v}, \label{linear-toner-tu-eq-app1}
\\
\partial_t \delta \mathbf{v} + \lambda (\mathbf{v}_{\rm ss} \cdot \nabla)\delta \mathbf{v} \! &=& \! -2\beta(\mathbf{v}_{\rm ss}\cdot \delta\mathbf{v})\mathbf{v}_{\rm ss} -c^2 \frac{\nabla \delta \rho}{\rho_{\rm ss}}.
\label{linear-toner-tu-eq-app2}
\end{eqnarray}
Finally, we neglect the second-order contributions of $\mathbf{v}_{\rm ss}$ and arrive at the linearized equation~(3) in the main text. 

We can also derive the Schr\"{o}dinger-like equation from Eqs.~\eqref{linear-toner-tu-eq-app1}, \eqref{linear-toner-tu-eq-app2} with a little algebra. Applying the convective derivative $\partial_t + \lambda (\mathbf{v}_{\rm ss}\cdot\nabla)$ to Eq.~\eqref{linear-toner-tu-eq-app1}, we obtain
\begin{equation}
(\partial_t + \lambda (\mathbf{v}_{\rm ss}\cdot\nabla))(\partial_t + (\mathbf{v}_{\rm ss}\cdot\nabla)) \delta \rho = c^2\nabla^2 \delta\rho.
\end{equation}
Assuming an oscillating solution $\delta \rho(\mathbf{r},t) = \delta \tilde{\rho}(\mathbf{r}) e^{i\omega t}$, the equation takes the form as
\begin{equation}
[c^2 \nabla^2 + \omega^2 -i\omega(\lambda+1)\mathbf{v}_{\rm ss}\cdot \nabla] \delta \tilde{\rho} = 0,
\end{equation}
where we assume that $|\mathbf{v}_{\rm ss}|$ is small and neglect its second-order terms. This equation is equivalent to the Schr\"{o}dinger-like equation~(5) in the main text aside the second-order contributions of $|\mathbf{v}_{\rm ss}|$.

\subsection{Deformation of the effective Hamiltonian in the basis appropriate for the band calculations}
To numerically obtain the band structure in the continuum space, we must in practice discretize the effective Hamiltonian with respect to spatial degrees of freedom. Specifically, we approximate the continuum space by a discretized triangular lattice and consider quantities on each lattice point as values of density  and velocity fields. Also, the derivatives $\partial_{x,y}$ must be converted into the difference between neighboring sites \cite{Smith1985}. The way of this conversion is not unique and can lead to  numerical errors; a naive discretization of the effective Hamiltonian can fail to provide the correct band structure. Since our system is symmetric under $\pi/3$ rotation, the band structure should also reflect that symmetry. However, one cannot symmetrize a pair of linear combinations of the derivatives $\partial_{x,y}$ under $\pi/3$ rotation. This is the main reason why the calculated band structures can in practice break the symmetry or have substantial numerical errors due to large wavenumber components. 

To solve this problem, we deform the effective Hamiltonian without affecting the topology of the system. First, we add the redundant degree of freedom $\delta \tilde{v}_r$ and formally rewrite the linearized equation~(3) in the main text as follows:
\begin{equation}
 \mathcal{H}'' \left(
  \begin{array}{c}
   \delta \tilde{\rho} \\
   \delta \tilde{v}_x \\
   \delta \tilde{v}_y \\
   \delta \tilde{v}_r
  \end{array}
  \right) =\omega
  \left(
  \begin{array}{c}
   \delta \tilde{\rho} \\
   \delta \tilde{v}_x \\
   \delta \tilde{v}_y \\
   \delta \tilde{v}_r
  \end{array}
 \right),
\end{equation}
where we denote ${\cal H}''$ as
\begin{equation}
  \mathcal{H}'' = \left(
  \begin{array}{cccc}
   -i\mathbf{v}_{\rm ss} \cdot \nabla & -i\partial_x & -i\partial_y & 0 \\
   -i\partial_x & -i\lambda\mathbf{v}_{\rm ss} \cdot \nabla & 0 & 0 \\
   -i\partial_y & 0 & -i\lambda\mathbf{v}_{\rm ss} \cdot \nabla & 0 \\
   0 & 0 & 0 & 0
  \end{array}
  \right).
\end{equation}
This deformation only adds trivial eigenstates with zero eigenvalues. We next consider the unitary matrix
\begin{equation}
  U = \left(
  \begin{array}{cccc}
   1 & 0 & 0 & 0 \\
   0 & \frac{2}{\sqrt{6}} & 0 & \frac{1}{\sqrt{3}} \\
   0 & -\frac{1}{\sqrt{6}} & \frac{1}{\sqrt{2}} & \frac{1}{\sqrt{3}} \\
   0 & -\frac{1}{\sqrt{6}} & -\frac{1}{\sqrt{2}} & \frac{1}{\sqrt{3}} 
  \end{array}
  \right),
\end{equation}
and use it to transform $\mathcal{H}''$ into the following form:
\begin{eqnarray}
  &{}& \mathcal{H}' = U \mathcal{H}'' U^{-1} \nonumber\\
  &=& -i \left(
  \begin{array}{cccc}
   \mathbf{v}_{\rm ss} \cdot \nabla & \frac{2}{\sqrt{6}} \partial_1 & \frac{2}{\sqrt{6}} \partial_2 & \frac{2}{\sqrt{6}} \partial_3 \\
   \frac{2}{\sqrt{6}} \partial_1 & \frac{2\lambda}{3}\mathbf{v}_{\rm ss} \cdot \nabla & -\frac{\lambda}{3}\mathbf{v}_{\rm ss} \cdot \nabla & -\frac{\lambda}{3}\mathbf{v}_{\rm ss} \cdot \nabla \\
   \frac{2}{\sqrt{6}} \partial_2 & -\frac{\lambda}{3}\mathbf{v}_{\rm ss} \cdot \nabla & \frac{2\lambda}{3}\mathbf{v}_{\rm ss} \cdot \nabla & -\frac{\lambda}{3}\mathbf{v}_{\rm ss} \cdot \nabla \\
   \frac{2}{\sqrt{6}} \partial_3 & -\frac{\lambda}{3}\mathbf{v}_{\rm ss} \cdot \nabla & -\frac{\lambda}{3}\mathbf{v}_{\rm ss} \cdot \nabla & \frac{2\lambda}{3}\mathbf{v}_{\rm ss} \cdot \nabla
  \end{array}
  \right). \nonumber\\
\end{eqnarray}
Note that the eigenvalues are unchanged and the eigenvectors are only globally transformed by $U$. Thus, these transformations do not alter the topology of the system. However, this new basis now naturally reflects the underlying symmetry of kagome lattice and, in practice, allows one to accurately calculate the band structure with much less numerical errors than the naive discretization in the basis of $x,y$ directions.   
To further improve the numerical accuracy, we also implement a hybrid discretization of the derivatives. More specifically, to convert the derivatives into the discretized form, we use a forward difference, a backward difference and a central difference for the derivatives $\partial_{1,2,3}$ in the first row of the effective Hamiltonian $\mathcal{H}''$, the derivatives $\partial_{1,2,3}$ in the first column of the effective Hamiltonian $\mathcal{H}''$, and the derivative $\nabla$, respectively. This conversion sustains the Hermiticity of the effective Hamiltonian and can remove  substantial numerical errors that can contribute from large wavenumber components.

\begin{figure}[t]
  \includegraphics[width=70mm]{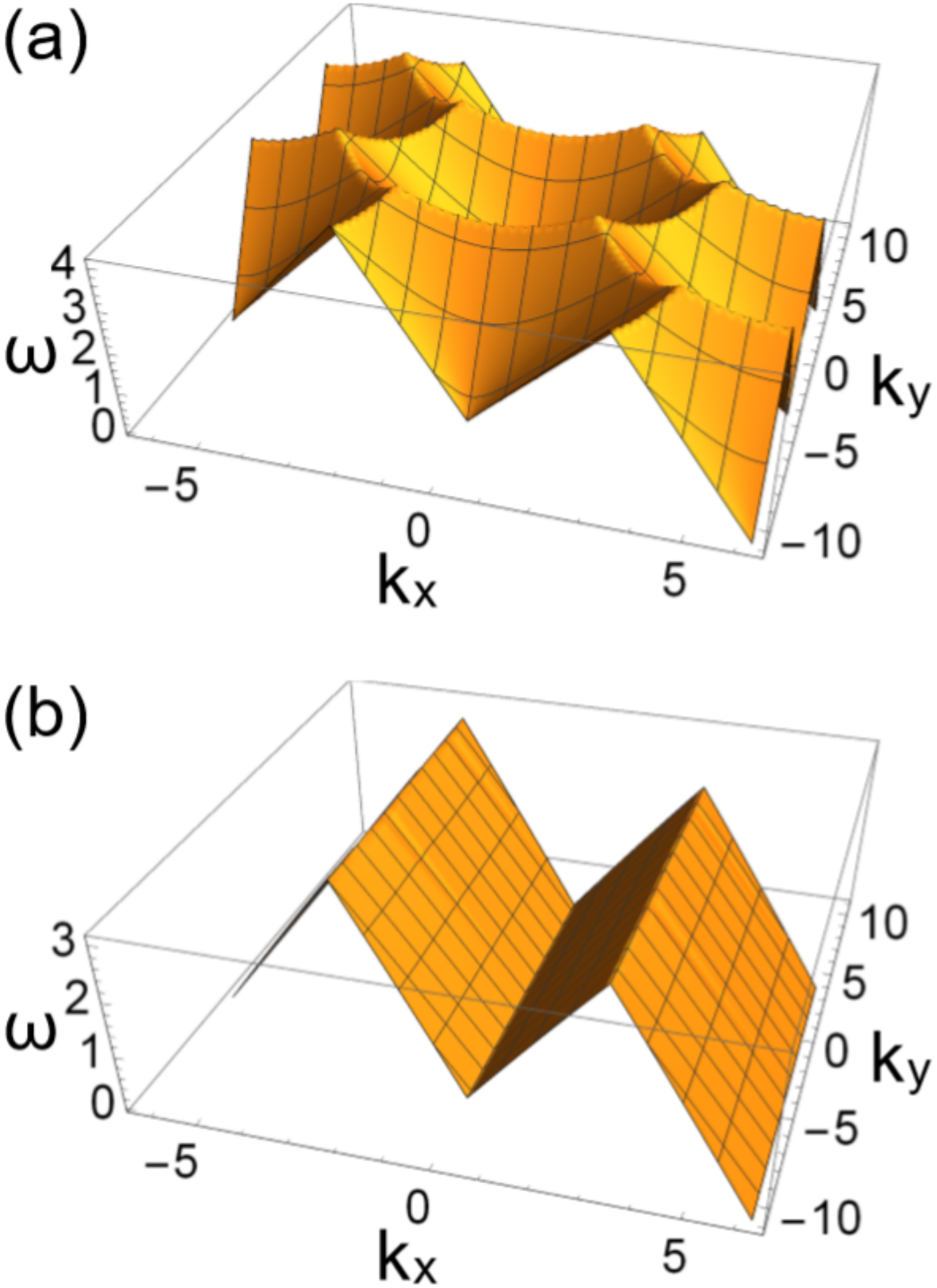}
\caption{\label{fig:unphysical-band} (a) The correct band structure obtained for the continuum system discretized with the redundant degrees of freedom. (b) Unphysical band structure obtained by the numerical calculations without the redundant degrees of freedom.}
\end{figure}

We compare the bulk band structures for the unordered state, $|\mathbf{v}_{\rm ss}|=0$, obtained by the calculations with and without the redundant degree of freedom. As shown in Supplementary Figure~\ref{fig:unphysical-band}(a), with the proper discretization of the effective Hamiltonian with the redundant degree of freedom, we can obtain the linear dispersion symmetric under $\pi/3$ rotation. On the contrary, Supplementary Figure~\ref{fig:unphysical-band}(b) demonstrates that the bulk band without the redundant degree of freedom shows the incorrect dispersion that is  independent of $k_y$. We can also obtain the analytical expressions of the dispersion relation for the unordered state. The correct dispersion is $\omega=0,\pm|\mathbf{k}|$ while we obtain the following dispersion from the calculation without the redundant degree of freedom
\begin{equation}
\omega = 0, M\sqrt{\sin^2 \left( \frac{k_x+2\pi n}{M} \right) + \frac{1}{3} \left[ \sin\left(\frac{k_x/2+(\sqrt{3}/2)k_y+2\pi m}{M}\right) + \sin\left( \frac{-k_x/2+(\sqrt{3}/2)k_y+2\pi (m-n)}{M} \right) \right]},
\end{equation}
where $M$ is the mesh number for the numerical calculation and $n,m=0,1,\cdots,M-1$. Considering $m=0$, $n=M/2$, $k_x=0$, the frequency is independent of $k_y$ and always $0$. With the other values of $k_x$, the frequency can be also almost independent of $k_y$ if we choose a certain integer $n$.  Even if we consider nonzero $|\mathbf{v}_{\rm ss}|$, these unphysical bands still remain and affect the topology of the system. Therefore we have to introduce the redundant degree of freedom.

\subsection{Bulk bands and edge dispersions}
In our model, we find that some edge bands exist deep inside the bulk bands. 
In practice, it is difficult to graphically demonstrate that those edge bands actually connect gapped bulk bands. This is because the bulk bands are gapped but not ``fully gapped" in the sense that bulk bands and an edge mode can have the same frequency but at different $k_x$. As a consequence, in the open-boundary spectrum $E(k_y)$, the bands appear to overlap even if the corresponding bulk bands are actually gapped.

\begin{figure}[t]
  \includegraphics[width=120mm]{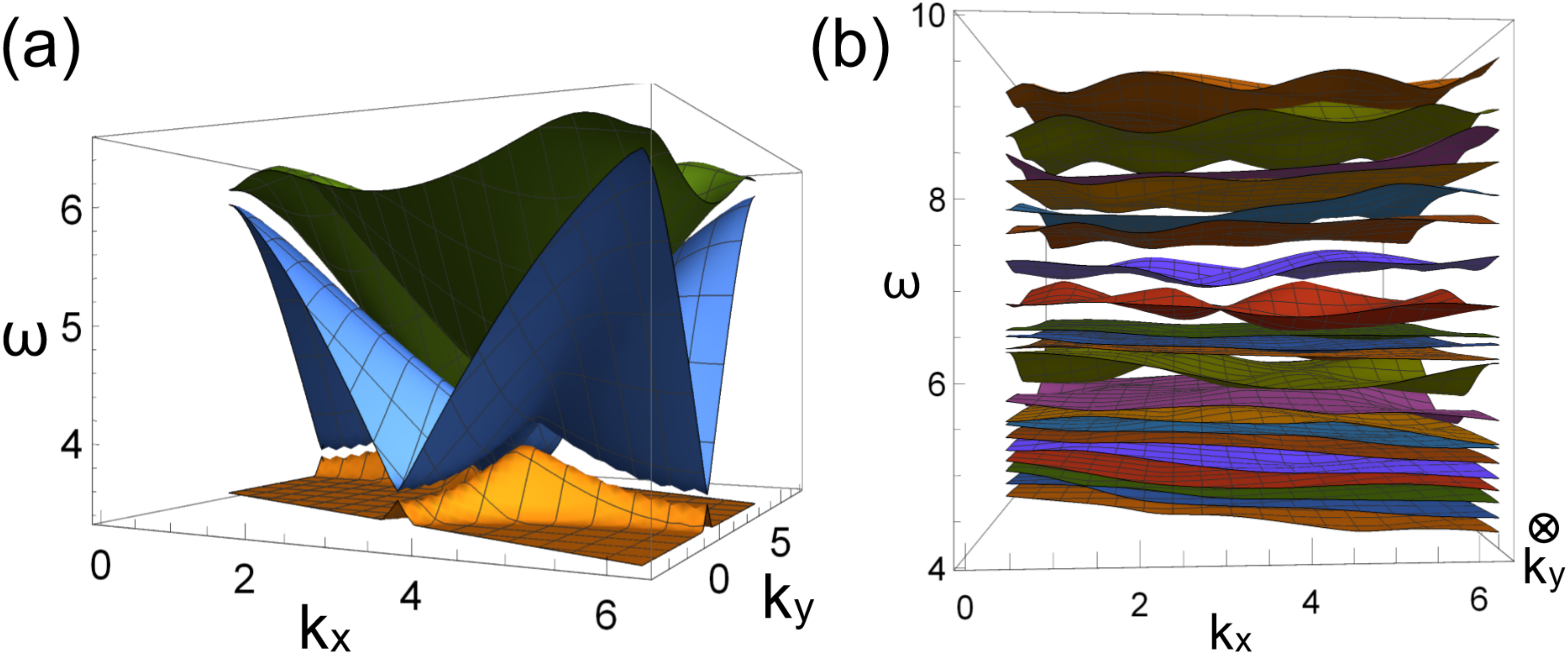}
\caption{\label{fig:bulkband} Bulk band structures under the periodic boundary conditions (a) around $\omega = 4.2$ for $|\tilde{\mathbf{v}}_{\rm ss}|=0.2$ and (b) around $\omega = 7.6$ for $|\tilde{\mathbf{v}}_{\rm ss}|=0.5$.}
\end{figure}

To explain this point more explicitly, we calculate the bulk dispersion (under the periodic boundary conditions) for the parameters we use in the main text. Supplementary Figure~\ref{fig:bulkband}(a) shows the bulk dispersion relation around the frequency $\omega = 4.2$ for $|\tilde{\mathbf{v}}_{\rm ss}|=0.2$, which corresponds to the gap in Fig. 2(b) in the main text. We can confirm that the three bands are separated from each other and thus are gapped. However, they are not ``fully gapped" in the above sense, i.e., no flat planes parallel to the $k_x$-$k_y$ plane exist without crossing the three bands. On the other hand, the bulk bands around the frequency $\omega = 7.6$ for $|\tilde{\mathbf{v}}_{\rm ss}|=0.5$ (corresponding to Fig.~4(b) in the main text) are ``fully gapped" as shown in Supplementary Figure~\ref{fig:bulkband}(b). Indeed, there exists a flat plane across the whole Brillouin zone without touching to the bulk bands (cf. the gray dashed horizontal line in Fig.~4(b) in the main text). This helps us to graphically demonstrate the existence of an edge dispersion connecting the bulk bands.

\end{document}